\documentclass[12pt]{article}
\usepackage[utf8]{inputenc}
\usepackage[export]{adjustbox}
\usepackage{authblk}
\usepackage{bbm}
\usepackage{pgfpages}
\usepackage{graphicx}
\usepackage{multicol}
\usepackage{csquotes}
\usepackage{multirow}
\usepackage{graphicx} 
\usepackage{booktabs} 
\usepackage{authblk}
\usepackage{caption}
\usepackage{subcaption}
\usepackage{stackengine}
\usepackage{amsmath}
\usepackage{comment}
\usepackage{hyperref}
\hypersetup{
    colorlinks=true,
    linkcolor=blue,
    filecolor=blue,      
    urlcolor=blue,
    citecolor=blue,
}
\usepackage[capposition=top]{floatrow}

\usepackage[spanish,english]{babel}

\usepackage{amssymb}
\usepackage{stackrel}
\usepackage{natbib} 

\usepackage[a4paper, margin=1in]{geometry}

\title{\textbf{TANK meets Diaz-Alejandro} \\ \large Household heterogeneity, non-homothetic preferences \& policy design}

\author[1]{Santiago Camara\footnote{Address: 2211 Campus Dr, Evanston, IL 60208. E-mail: santiagocamara2022@u.northwestern.edu . I would like to thank the Becker Friedman Institute and the Monetary and Fiscal History of Latin America research project for funding this project. I would also like to thank Lawrence Christiano, Giorgio Primiceri and Matthew Rognlie for their guidance. Francisco Roldan and Daniel Heymann provided helpful comments.}}

\affil[1]{Northwestern University \& Red-NIE}
\date{\today}

\begin{document}

\maketitle

\begin{abstract}
    
    This paper studies the role of households' heterogeneity in access to financial markets and the consumption of commodity goods in the transmission of foreign shocks. First, I use survey data from Uruguay to show that low income households have poor to no access to savings technology while spending a significant share of their income on commodity-based goods. Second, I construct a Two-Agent New Keynesian (TANK) small open economy model with two main features: (i) limited access to financial markets, and (ii) non-homothetic preferences over commodity goods. I show how these features shape aggregate dynamics and amplify foreign shocks. Additionally, I argue that these features introduce a redistribution channel for monetary policy and a rationale for \textit{fear-of-floating} exchange rate regimes. Lastly, I study the design of optimal policy regimes and find that households have opposing preferences a over monetary and fiscal rules.
    
    \medskip
    
\end{abstract}

\newpage
\section{Introduction} \label{sec:introduction}

The impact of foreign shocks on aggregate outcomes and inequality in small open economies is a classic question in international macroeconomics, starting with the seminal work of \cite{alejandro1963note}, \cite{fleming1962domestic} and \cite{mundell1963capital}. Furthermore, the design of the optimal policy mix in response to foreign shocks is still an open question.\footnote{For instance, \cite{camaraetal2021} study the cost and benefits of foreign exchange rate interventions. Also, \cite{garcia2015dealing} study optimal fiscal policies in response to commodity price shocks in the presence of heterogeneous households and learning-by-doing frictions. Also, \cite{cugat2019emerging} study optimal exchange rate policies in the presence of heterogeneous households in a two-sector economy.} Recent literature has argued that household heterogeneity plays a crucial role in shaping aggregate dynamics in response foreign shocks (see \cite{cugat2019emerging}, \cite{de2020household} and \cite{auclert2021exchange}). In this paper, I argue that accounting for households heterogeneity in access to financial markets and in consumption bundles is crucial for understanding both aggregate dynamics and the design of optimal policies in small open economies

To begin with, I use survey data on Uruguayan households to characterize key dimensions of household heterogeneity. First, I argue that households differ significantly in their access to savings technologies. In particular, more than half of the households do not have access to banking accounts, less than a third report having any savings and less than 10\% hold any financial assets. In addition, I show that low income households are less likely to have access to these financial tools. Second, I present evidence of Engel's law or non-homothetic preferences, i.e., low income households spend a significantly higher share of their income in nourishment than high income households. Consequently, low income households are less likely to smooth consumption across time and are more exposed to changes in the price of food.\footnote{Furthermore, I argue that for the case of Uruguay, the domestic price of food is significantly correlated with international food and agricultural commodity prices. In consequence, I argue that domestic households are exposed to changes in international food prices.}

Then, I construct a Two Agent New Keynesian (``TANK'') small open economy with two main features (i) limited access to financial markets across households (Ricardian and Hand-to-mouth); and (ii) non-homothetic preferences over commodity goods. Thus, Hand-to-mouth agents endogenously spend a higher share of their income in commodity goods than Ricardian households, especially during periods of low income. I show how aggregate and household specific dynamics in response to foreign interest rate and commodity price shocks depend on these novel model features. I argue that non-homothetic preferences exacerbate the impact of foreign shocks, particularly of Hand-to-mouth households which do not have access to consumption smoothing technologies.

Finally, following the description of the model I evaluate the design of optimal policy mixes. First, I argue that under this structural framework monetary policy affects households through a distributional channel. On the one hand, a monetary tightening reduces the consumption of Ricardian households through an intertemporal substitution channel and a reduction of economic output. On the other hand, a monetary tightening appreciates the nominal exchange rate, reducing the domestic price of commodity goods and thus increasing the consumption of Hand-to-mouth households. Additionally, I introduce a fiscal rule which depends on the commodity income. Ricardian and Hand-to-mouth households have opposing preferences over policy mixes. Lastly, I show that households have opposing preferences over monetary and fiscal policy regimes. A monetary policy rule which stabilizes the nominal exchange rate, and consequently the price of Hand-to-mouth's consumption bundle, provides a new rationale for Central Banks' \textit{fear-of-floating}.

\noindent
\textbf{Related literature.} This paper relates to three different strands of international macroeconomics. First, it relates to a literature which studies the sources of fluctuations in small open economies. Seminal papers in this literature are \cite{mendoza1991real}, \cite{neumeyer2005business} and \cite{garcia2012real} (emphasizing the role of foreign interest rate shocks), \cite{medina2007copper} (emphasizing the role commodity price shocks) and \cite{gali2005monetary} (in a New Keynesian set up). These papers study how shocks to international interest rates and to commodity price impact small open economies. Additionally, this literature emphasis that these foreign shocks explain a significant share of aggregate fluctuations. For instance, \cite{drechsel2018commodity} estimate a medium scale DSGE model for Argentina and show that commodity price shocks explain close to 40\% of fluctuations in GDP and consumption. This paper contributes to this literature by introducing a structural framework where households have non-homothetic preferences over the consumption of commodity goods. Consequently, commodity goods explain a significant share of both consumption and export bundles. I argue that these features exacerbate the impact of foreign shocks.\footnote{Other papers have introduced non-homothetic preferences to analyze foreign or international shocks. For instance, \cite{do2017trade} study the role of non-homothetic preferences in the design of optimal trade policies. More recently, \cite{rojas2020non} study the role of non-homothetic preferences during Sudden Stop crises in a tradable and non-tradable setup. This paper differs from the latter paper as it focuses on the role of commodity goods and the design of optimal monetary and fiscal policies.}

Second, this paper relates to a literature which studies the impact of agent heterogeneity in small open economies in propagating shocks. For example, \cite{cravino2017distributional} study the distributional effects of large exchange rate depreciations. Also, \cite{de2020household} develop a Heterogeneous Agent New Keynesian (HANK) model to study the role of household heterogeneity during a current account reversal episode. More recently, \cite{auclert2021exchange} use a HANK model to quantify the real income channel, i.e., the fall on real income due to a rise in import prices from a depreciation. More closely to the present paper, \cite{cugat2019emerging} argues that household heterogeneity played a significant role during the 1995 Mexican financial crisis. To quantify the role of this heterogeneity she introduces Ricardian and Hand-to-mouth households and sector specific labor to a New Keynesian set up. This paper contributes to this literature by focusing on households' heterogeneous exposure to commodity consumption through non-homothetic preferences. I show that this dimension of household heterogeneity amplifies the impact of foreign shocks, but even more so when combined with limited access to financial markets. 

Third, this paper relates to a literature which analyzes the design of optimal policies in small open economies. The seminal papers in this literature are \cite{parrado2002optimal} and \cite{gali2005monetary} which focus on interest rate rules and \cite{camaraetal2021} which focus on exchange rate interventions. However, these papers focused on representative agent models. For two-agent economies, the work of \cite{garcia2015dealing} and \cite{cugat2019emerging} stand out. The former paper studies optimal fiscal and macro prudential policies to deal with the ``Dutch disease'' generated by commodity price shocks in a small open economy populated by Ricardian and Hand-to-mouth households. The latter paper studies optimal monetary policies in an economy populated with Ricardian and Hand-to-mouth households which are also sector-specific (tradable and non-tradable). In line with these two papers, I find that households differ in their preferences over policies mixes. This paper contributes to the literature by providing a novel rationale for the ``fear of floating'' phenomenon in which Central Banks prefer less flexible exchange rate regimes.\footnote{For details on the discussion of ``fear of floating'' see \cite{calvo2002fear}.} Hand-to-mouth households prefer Taylor rule regimes which react to exchange rate depreciations as it stabilizes the domestic price of commodity goods (the significant component of their consumption bundle). The welfare gains for less flexible exchange rate regimes is increasing in the degree of non-homothetic preferences.

\noindent
\textbf{Organization.} This paper consists of \ref{sec:conclusion} sections, starting with this introduction. Section \ref{sec:data} describes the data sets used in the paper and presents micro level stylized facts. Section \ref{sec:model} presents the key model features and Section \ref{sec:dynamics} describes how these features matter for the transmission and amplification of foreign shocks. Section \ref{sec:optimal_policy} studies the design of optimal policy mixes. Section \ref{sec:conclusion} concludes. 

\section{Data Description \& Households Stylized Facts} \label{sec:data}

In this section of the paper I describe the household level data and present stylized facts which characterize key patterns of income, savings and consumption. I argue that two key dimensions of household heterogeneity are the access to savings technologies and the importance of commodity based goods in consumption baskets. 

To start, I describe my micro level data set. I use survey data on Uruguayan households coming from the "\textit{Encuesta Financiera de los Hogares Uruguayos}". This survey was carried out by the \textit{Facultad de Ciencias Sociales} of the \textit{Universidad de la Republica} in three stages: 2012, 2014 and 2017. The design of the sample of this survey is based on Uruguay's 2011 population census. The survey collects cross-sectional socio-demographic and economic-financial information on households, revealing in detail the possession, composition and value of assets and liabilities, income and expenditure, as well as proxy variables to measure access to financial markets and use of means of payment. In total, the survey presents data on 3490 households.

Next, I turn to introducing stylized facts on the Uruguayan households' patterns of savings and consumption. I focus on describing households' access to different savings technologies. In argue that low income households have a lower probability of holding savings, owning financial assets or owing financial debt. Additionally, I show that consumption patterns exhibit Engel's law, i.e., the share of income spent on food and nourishment is decreasing on income.

\noindent
\textbf{Fact 1:} \textit{The share of households which report savings or access to financial tools is low, approximately 20\% and 36\% respectively. Additionally, the probability of having savings or access to financial tools is strongly correlated with income.}

To start, Fact 1 argues that in the cross-section of Uruguayan households there is significant heterogeneity in access to saving technologies. 
\begin{figure}[ht]
    \centering
    \caption{Households' Access to Saving Technologies \\ Binary Response Indicators }
    \label{fig:Fact1}
    \includegraphics[width=16cm,height=10cm]{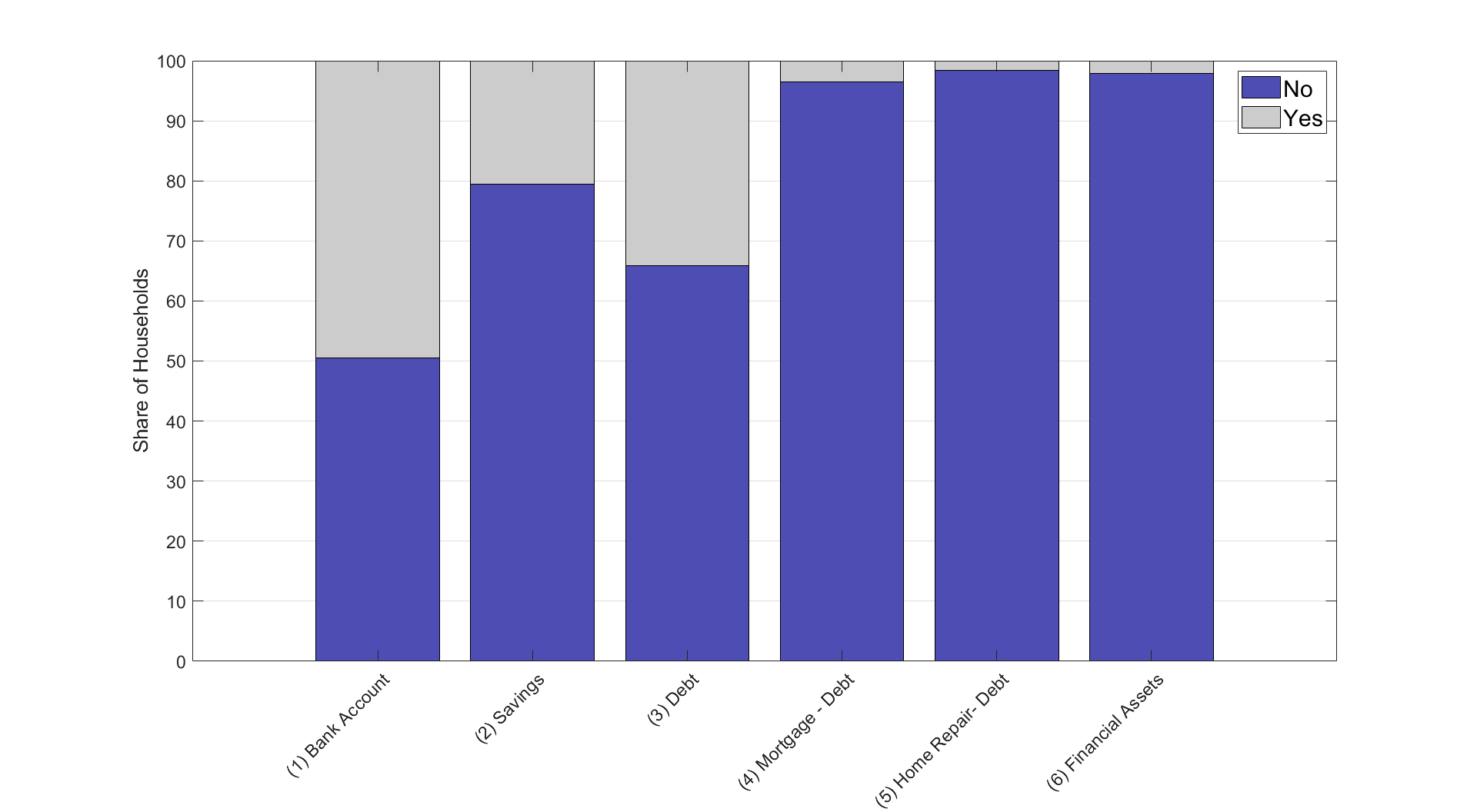}
\end{figure}
First, Figure \ref{fig:Fact1} shows the households' response to binary survey questions on the access to savings technology. Column (1) shows that only half the households in the sample have at least one member having a bank account (savings or checking account). Column (2) shows that only 20\% of households answer ``Yes'' to having any kind of savings. Column (3) shows that close to 34\% of households have some kind of debt. Columns (4) and (5) show that households' debt is not primarily explained by housing mortgages or debt to carry out home reparations. Column (6) shows that only 2\% of surveyed households answer ``Yes'' to owning financial assets. Overall, the main message this figure conveys is that a significant share of households in the survey have no access to saving technologies. This is line, with a previous literature which has found that significant shares of households in Latin America do not have access to financial markets (see \cite{de2004inequality}, for instance).

Next, I show that the probability of a household reporting savings, debts or holding financial assets is highly correlated with income. To do so, I carry out a cross-sectional regression analysis of the form
\begin{align} \label{eq:regression_prob_savings}
    \mathbbm{1}\left[ Y_i = 1 \right] = \beta \ln \text{Income}_i + \gamma X_{i} + \epsilon_{i}
\end{align}
where $Y_i$ are binary variables in response to the following set of questions: (i) \textit{Any member of the household has a bank account (savings or checking account)?}, (ii)  \textit{Household has any kind of savings?}, (iii) \textit{Household has any kind of debt?}, (iv) \textit{Household has any kind of financial asset?}, (v) \textit{In the last 5 years, has any member of the household had a credit denied?} which take the value of one if the answer to the question is ``Yes''. Parameter $\beta$ represents the relationship between these binary variables and log total household income, $X_i$ is a set of household controls which include main income earner's educational level and age.\footnote{The survey is designed to be answered by the household's main income earner. Consequently, I assume that the person who answers the survey or ``Person of Reference'' is the main income earner. As a robustness check I carry out empirical exercises such as the one in Equation \ref{eq:regression_prob_savings} by using the average educational level and age of the ``Person of reference'' and his/her partner if they have one. Results are robust.}\footnote{Educational level is defined as the last education level the person started. The different educational levels are: never attended any educational level, elementary school, middle school, high-school or equivalent, technical secondary degree, tertiary degree, professorship degree, non-college tertiary degree, college degree, postgraduate degree.}\footnote{Age is controlled through a quadratic function to capture a life-cycle component.} Table \ref{tab:reg_fact1} presents the results of estimating Equation \ref{eq:regression_prob_savings} using a linear probability model. 
\begin{table}[ht]
    \centering
    \caption{Households' Income \& Saving Technologies}
    \label{tab:reg_fact1}
\begin{tabular}{lccccc}
& Bank Acc. & Savings & Debt & Fin. Assets & Credit Denied \\
& (1) & (2) & (3) & (4) & (5) \\ \hline \hline
 &  &  &  &  \\
$\ln \text{Income}_i$ & 0.208*** & 0.142*** & -0.00817 & 0.0211*** & -0.0137** \\
 & (0.0102) & (0.00918) & (0.0119) & (0.00366) & (0.00550) \\
 &  &  &  &  \\
HH. Controls  & YES & YES & YES & YES & YES  \\
 &  &  &  &  &  \\
Observations\footnote{While 3,490 households participate in the survey, 3 households do not report valid ages. This explains the lower number of observations in these regressions.} & 3,487 & 3,487 & 3,487 & 3,487 & 3,487 \\
 R-squared & 0.286 & 0.212 & 0.010 & 0.058 & 0.018  \\ \hline \hline
\multicolumn{6}{c}{ Robust standard errors in parentheses} \\
\multicolumn{6}{c}{ *** p$<$0.01, ** p$<$0.05, * p$<$0.1} \\
\end{tabular}
\end{table}
Except the binary variable denoting debt holdings, all other $Y_i$ variables are highly correlated with the households' level of income. Consequently, the evidence seems to suggest that low income households have limited access to savings technologies. This result is in line with several papers in the literature.\footnote{For instance,  \cite{hong2020emerging} compute marginal propensities to consume for Peru and show that households are significantly more credit constrained than American households.}\footnote{Given the lack of a time-series dimension to the dataset is impossible to test whether the lack of access to savings, debt or financial assets by low income households is cyclical (e.g. a household going through a temporary unemployment spell) or permanent (low educated households being systematically rationed out of financial markets). Furthermore, an incomplete market model with uninsured idiosyncratic risk a la \cite{aiyagari1994uninsured} predicts a positive correlation between income and household savings under standard parametrizations.}

\noindent
\textbf{Fact 2:} Low income households spend a greater share of income in food and beverages than high income households.

The second fact highlights differences in the consumption basket of low and high income households. In particular, low income households spend a significant share of their income on nourishment.\footnote{The survey question asks: ``\textit{Approximately how much does the household spend on nourishment?}(alimentaci\'on).''} This is a long standing stylized fact in economics usually referred to as Engel's law.
\begin{figure}[ht]
    \centering
    \caption{Share of Income Spent on Nourishment by Income Level}
    \label{fig:Fact2}
    \includegraphics[scale=0.75]{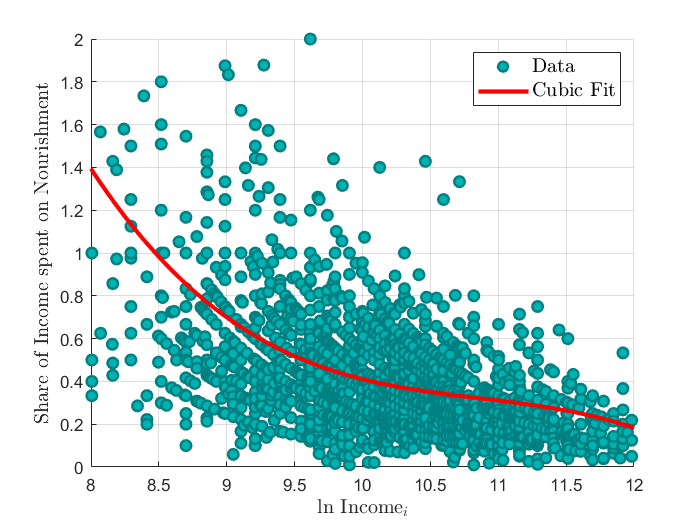}
    \floatfoot{\textbf{Note}: The red line is computed as the cubic fit which produces the best fit (in a least-squares sense).}
\end{figure}
Figure \ref{fig:Fact2} presents the relationship between the share of expenditure on nourishment and log income. This graph presents strong evidence in favor of Engel's law.\footnote{In the left tail of the distribution, several households report spending more on nourishment than their total income. This could be driven by measurement error or households using accumulated savings or relying on formal or informal credit/debt during periods of low income/unemployment spells.} This is, there seems to be a non-homothetic relationship between income and expenditure in nourishment. On the one hand the average share of income spent on nourishment for households under the median reported income is 55,2\%.  On the other hand, households above the median on average spend 32.3\% of their income on nourishment. In consequence, low income households are more exposed to changes in food prices than high income households. This is of significant importance as Uruguay's export bundle is highly skewed toward agricultural or food commodity products.\footnote{For more evidence on this, see Appendix \ref{subsec:appendix_firm_findings}.}

Finally, I present evidence which connect ``Fact 1'' and ``Fact 2'' by arguing that households with lower access to savings technologies spend a higher share of their income on nourishment.
\begin{table}[ht]
    \centering
    \begin{tabular}{l|c c c c c}
     & \multicolumn{5}{c}{Question: Household has access to or holds \_\_\_ ?} \\
      & Bank Acc.  &  Savings &  Debt &  Fin. Assets & Credit Denied  \\
     \underline{Answer} & (1)  &  (2) &  (3) &  (4) & (5)  \\ \hline \hline
     No      & 50.8\% & 47.2\% & 42.6\% & 44.1\% & 43.4\% \\
     Yes     & 36.5\% & 30.1\% & 45.8\% & 27.4\% & 48.9\% \\
     \\
     t-test. & 11.36  & 11.05  & -2.38  & 3.75   & -1.87 \\ \hline \hline
    \floatfoot{\small \textbf{Note:} The two entries of each column of the table present the average share of income spent on Nourishment according to the answer to the specific survey question. The ``t-test'' is computed carrying out a standard ``t-test'' which test whether the group specific average share of income spent on nourishment is different.}
    \end{tabular}
    \caption{Share of income spent on nourishment \\ by access to saving technologies }
    \label{tab:Fact2}
\end{table}
Table \ref{tab:Fact2} presents the average household share of income spent on nourishment according to whether they have access to the different saving technologies studied above. Columns (1), (2) and (4) show that on average, households which have access to bank accounts, report savings or report holding financial assets spend a significantly lower share of income on nourishment than households which do not have access to these savings technologies (between 30\% and 60\% lower). Columns (3) and (5) show that on average, households which either report debt or having a credit denied in the last 5 years spend significantly more on nourishment than household which do not report debt or have not had credit requests denied. The last row of Table \ref{tab:Fact2} carries out a ``t-test'' to show that these group means are statistically different from each other.

In summary, this section used survey data to highlight two dimensions of household heterogeneity. To do so, I presented two stylized facts on households' savings and consumption patterns emerging from cross-sectional data. Fact 1 argues that a significant share of households in Uruguay do not have access to saving technologies. To this end, I show that a notable share of households report no savings or hold no financial assets. Furthermore, I argue that the probability of reporting access to savings technologies is positively correlated with income. Fact 2 argue that relatively low income households spend a significantly higher share of their income on nourishment than relatively higher income households. These stylized facts are the cornerstone of the theoretical framework constructed in \cite{alejandro1963note}. To test the aggregate implications of these dimensions of household heterogeneity, in the next section I construct a macroeconomic model inspired on the stylized facts introduced in these facts.

\section{Model} \label{sec:model}

This section of the paper presents a structural Two Agent New Keynesian model of a small open economy. The key novel features of this model are: Ricardian and Hand-to-mouth households and non-homothetic preferences over the consumption of a commodity good. Section \ref{subsec:model_hh} describes the two households' optimality conditions. Section \ref{subsec:exports} describes the exporting of both commodity and non-commodity goods. The rest of the model is relatively standard and leave the details for Appendix \ref{sec:appendix_model}.

\subsection{Households \& Non-Homotheticity Preferences} \label{subsec:model_hh}

The economy considered is populated by two type of households: Ricardian ("$R$") households, which have access to different savings technologies, and Hand-to-Mouth ("$H$") households which can not save and consequently consume the totality of their income. Parameter $\lambda^R \in \left(0,1 \right)$ represents the share of Ricardian households in this economy. 

\noindent
\textbf{Ricardian households.} This households consume, work, save in domestic and foreign bonds, accumulate capital. Their life-time utility is given by
\begin{equation*}
    \mathcal{U}^R = \mathbb{E}_0 \sum^{\infty}_{t=0} \beta^{t}\left[ \log \left(C^R_t \right) - \chi_R \frac{\left(l^R_t\right)^{1+\varphi}}{1+\varphi} + \mu_t h_t \left(\frac{S_t D^{*}_t}{P^{R,C}_t}\right) \right]
\end{equation*}
where $C^R_t$ and $l^R_t$ represent this households' consumption and hours worked respectively; $\beta$ is the discount factor, $\varphi$ is the inverse of the Frisch labor elasticity and parameter $\chi_R$ alters the dis-utility of labor and is used to match steady state hours worked; $S_t$ is the nominal exchange rate, $P^{R,C}_t$ is the Ricardian household consumption price index and $D^{*}_t$ represents the household holdings of foreign assets. Function $\mu_t h_t (.)$, generates a non-pecuniary externality with respect to their holdings of foreign denominated bonds, which induce independence of the deterministic steady state from initial conditions (see \cite{schmitt2003closing}). The functional form for function $h_t$ is given by
\begin{equation} \label{eq:h_function}
    h_t \left( \frac{S_t D^{*}_t}{P^{R,C}_t}  \right) = -\frac{1}{2} \gamma \left(\frac{S_t D^{*}_t}{P^{R,C}_t} - Z_t \Upsilon_t \right)^2
\end{equation}
where $\Upsilon_t$ represents the households preferred target over it's holdings of foreign assets.\footnote{The term $\mu_t = 1 / Z^2_t$ is introduced to accurately deflate the variables inside the parenthesis in Equation \ref{eq:h_function}. More details on the introduction of long run growth and the deflation of variables are presented in Section \ref{subsec:model_firms}.} 

The Ricardian household's budget constraint is given by
\begin{align*}
    P^{R,C}_t C^R_t + P^k_t \left[K_t - \left(1-\delta^k\right)K_{t-1}\right] + S_t D^*_t + D_t \\
        \qquad \qquad \qquad \leq S_t R^{*}_{t-1}D^{*}_{t-1} + W_t l^R_t + R_{d,t-1}D_{t-1} +  \Omega^R_t 
\end{align*}
where $K_t$ represents the stock of capital, $P^k_t$ is the price of capital, $D_t$ are holdings of domestic asset, $R_{d,t}$ and $R^{*}_{t}$ are the interest rate on domestic and foreign assets, $W_t$ is workers' wage rate and $\Omega^R_t$ represents any profits from firms, commodity production and or net lump sum transfers from the government. Household's optimality conditions can be characterized by its intra-temporal labor choice and three Euler equations with respect to capital, domestic asset and foreign assets
\begin{align}
    \frac{W_t}{P^{R,C}_t} &= C^R_t \chi_R \left(l^R_t\right)^{\varphi} \\
    v_t &= \beta \mathbb{E}_t \left[v_{t+1} R^{k}_{t+1} \right] \\
    v_t &= \beta \mathbb{E}_t \left[ v_{t+1} R_{d,t} \right] \\
    \mu_t h^{'}_t \left(\frac{S_t D^{*}_t}{P^{R,C}_t}\right) \frac{S_t}{P^{R,C}_t} &= v_t S_t - \mathbb{E}_t \left[\beta v_{t+1} S_{t+1} R^{*}_{t} \right]
\end{align}
where $v_t$ is the Lagrange multiplier in the household's time $t$ budget constraint and $R^k_{t+1}$ is the rate of return on capital given by
\begin{equation*}
    R^k_t \equiv \left[ \frac{r_{t+1} + \left(1-\delta^k\right)P^k_{t+1}}{P^k_{t}} \right]
\end{equation*}
where $r_{t+1}$ is the real rental rate of capital. 

\noindent
\textbf{Hand-to-mouth Households.} The remaining fraction $1-\lambda^R$ of households consume and work, but are completely excluded from both financial and physical capital markets. This household's lifetime utility is described by
\begin{equation*}
    \mathcal{U}^{H} = \mathbb{E}_0 \sum^{\infty}_{t=0} \beta^t \left[ \log \left(C^{H}_t \right) - \chi_H \frac{\left(l^H_t\right)^{1+\varphi^H}}{1+\varphi^H}  \right]
\end{equation*}
where $C^{H}_t$ and $l^{H}_t$ represent consumption and hours worked respectively; $\beta$ is a discount factor, $\phi^H_t$ is the inverse of the Frisch labor elasticity. Parameter $\chi_H$ alters the dis-utility of labor and is used to match steady state hours worked across households. 

Hand-to-mouth households are not able to smooth their consumption and, thus, consume their total income every period. The representative hand-to-mouth household's budget constraint is given by
\begin{equation*}
    P^{H,C}_t C^{H}_t \leq W^H_t l^H_t + \Omega^H_t
\end{equation*}
where $P^{H,C}_t $ is the price of this households consumption bundle, $W^H_t$ is the nominal wage of the hand-to-mouth households, and $\Omega^H_t$ represent net taxes/transfers from the government. Hence, this household's optimality condition include the intra-temporal which relates labor and consumption
\begin{align}
    \frac{W_t}{P^{H,C}_t} &= C^H_t \chi_H \left(l^H_t\right)^{\varphi} 
\end{align}
but no intertemporal optimality conditions.

\noindent
\textbf{Non-Homothetic Preferences over Commodity Goods.} Both Ricardian and Hand-to-mouth households have non-homothetic preferences over the consumption of the commodity good. I assume households have the same preferences. Consequently, for simplicity of notation I suppress the household identifier from the rest of the section. The consumption bundle is comprised of a two tier nested aggregator. The first tier is modeled as a mix of Cobb-Douglas and Stone-Geary aggregator between commodity goods and non-commodity goods given by 
\begin{equation*} \label{eq:nh}
    C_t = \left(C^{Co}_t - \phi_{Co} \right)^{\alpha_{Co}} \left(C^{N}_t \right)^{1-\alpha_{Co}}
\end{equation*}
where $C^{Co}_t$ and $C^{N}_t$ represents households' consumption of commodity goods and non-commodity goods respectively. Parameter $\phi_{Co}$ governs the non-homotheticity of households' over commodity consumption.\footnote{The economy has trend growth, consequently, parameters $\phi^{j}_{Co}$ grows in steady state at the same rate of technology. For more details on the model see Appendix \ref{sec:appendix_model}.} This aggregator gives rise to following demand schedules
\begin{align}
    C^{Co}_t &= \frac{\alpha_{Co} C_t}{S_t P^{Co}_t \left( \frac{\alpha_{Co}}{S_t P^{Co}_t} \right)^{\alpha_{Co}} \left( \frac{1-\alpha_{Co}}{P^{N}_t} \right)^{1-\alpha_{Co}} } + \phi_{Co} \\
    C^{N}_t &= \frac{\left(1-\alpha_{Co}\right) C_t}{P^{N}_t \left( \frac{\alpha_{Co}}{S_t P^{Co}_t} \right)^{\alpha_{Co}} \left( \frac{1-\alpha_{Co}}{P^{N}_t} \right)^{1-\alpha_{Co}}}
\end{align}
where $S_t P^{Co}_t$ and $P^N_t$ are the domestic currency price of the commodity good and the domestic currency price of the non-commodity aggregator (further described below). The presence of non-homothetic preferences implies that consumers set aside a subsistence level of commodity goods, $S_t P^{Co}_t \phi_{Co,t}$, and allocate the remaining budget in proportion to parameter $\alpha_{Co}$. Consequently, the consumption of commodity goods exhibits Engel's law, i.e., relatively poor households spend a greater proportion of their income on food. To better understand the role of parameters $\phi_{Co}$ I compute the share of expenditure for the commodity and non-commodity goods for different values.
\begin{figure}[ht]
    \centering
    \caption{Expenditure Shares according to $\phi_{Co,t}$}
    \label{fig:nh_phiCo}
    \includegraphics[scale=0.75]{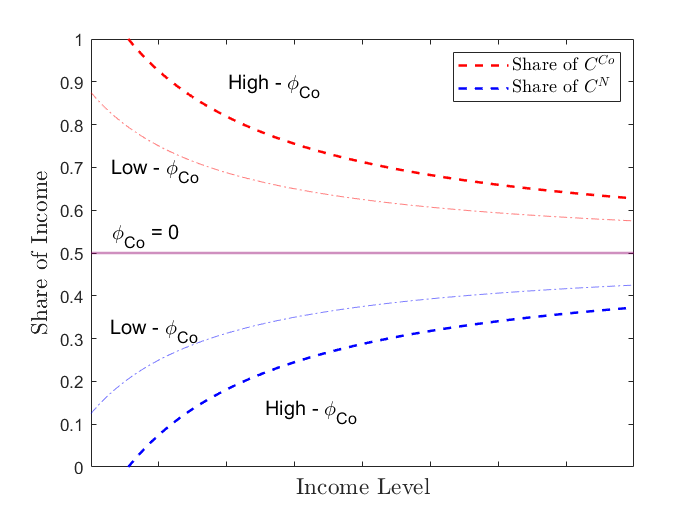}
\end{figure}
Figure \ref{fig:nh_phiCo} shows how expenditure shares on the commodity good change as income levels increase when $\alpha_{Co} = 0.5$. The solid magenta line shows that when $\phi_{Co} = 0$, the preferences presented in Equation \ref{eq:nh} collapse into the standard Cobb Douglas preferences, i.e., the expenditure share of the commodity good is equal to $\alpha_{Co}$. The dash and dot lines show the expenditure shares for the commodity (in red) and the non-commodity good (in blue) for a low but positive value of $\phi_{Co}$. Note that for low income levels the expenditure share in the commodity goods increase significantly (consequently the expenditure share in the non-commodity good decreases sharply). The skewness of expenditure share increases as $\phi_{Co}$ increases, as can be seen in the red and blue dashed lines in Figure \ref{fig:nh_phiCo}. 

Note that the domestic price of the commodity good is assumed to be equal to the international price $P^{Co}_t$ times the nominal exchange rate (which is the same assumption made for the domestic price of the imported goods in the rest of the model). This is not an innocuous assumption as it implies one-to-one exchange rate pass-through of the commodity good. In Appendix \ref{subsec:appendix_price_food} I study the dynamics of both international and Uruguay's domestic food indexes and present evidence in support of this model assumption.

Finally, the non-commodity good is itself a \textit{CES} aggregator of the homogeneous domestic  final good and the homogeneous imported good, given by
\begin{equation*}
    C^{N}_t = \left[ \left(\omega^{D}_c \right)^{\frac{1}{\eta_c}} \left(C^{D}_t\right)^{1-\frac{1}{\eta_c}} + \left(1-\omega^{D}_c\right)^{\frac{1}{\eta_c}} \left(C^{F}_t\right)^{1-\frac{1}{\eta_c}} \right]^{\frac{\eta_c}{\eta_c-1}}
\end{equation*}
where $C^{D}_t$ and $C^{F}_t$ represent the quantities consumed of the domestic and foreign final goods; $\varphi^{N}$ is a parameter that represents the relative importance of domestic goods in the non-commodity  aggregator and $\epsilon$ is a parameter governing the elasticity of substitution across the two goods. This framework yields standard demand schedules given by
\begin{align}
    C^{D}_t &= \omega^{D}_c \left(\frac{P^{N}_t}{P_t} \right)^{\eta_c} C^{N}_t \\
    C^{F}_t &= \left(1-\omega^{D}_c\right) \left(\frac{P^{N}_t}{S_t P^F_t} \right)^{\eta_c} C^{N}_t
\end{align}
where $P^{N}_t$ is the price of the non-commodity consumption bundle, and  $P^{F}_t$ is the price in foreign currency of the imported good.

\subsection{Exports} \label{subsec:exports}

This section describes the exporting side of the economy. In order to match the dynamics of a small open economy similar to those of Uruguay, exports are comprised of two distinct sub-sectors: a commodity sector and a non-commodity sector. On the one hand, the non-commodity good is produced using the domestic final good and exporters face a downward sloping demand curve which depend on a foreign demand shifter and the real exchange rate (in line with standard SOE-NK models). On the other hand, the commodity sector is characterized by stochastic processes which govern the international price of the good and the local endowment the commodity good.

\noindent
\textbf{Non-Commodity Exports.} There is a continuum of non-commodity exporting firms. The amount exported of non-commodity goods is denoted by $X_t$. Foreign demand for this good is given by 
\begin{equation} \label{eq:exports_noncom}
    X_t = \left(\frac{P^x_t}{P^f_t} \right)^{-\eta_f} Y^f_t
\end{equation}
where $P^x_t$ is the price in foreign currency of the exported good, $Y^f_t$ is a foreign demand shifter and $\eta_f$ is a parameter which governs demand elasticity. Foreign demand is exogenous to the domestic economy. In the above expression, $p_{t}^{x}$ is the non-commodity terms of trade: 
\begin{equation} \label{eq:terms_of_trade}
    p_{t}^{x}=\frac{P_{t}^{x}}{P_{t}^{f}}=\frac{P_{t}}{S_{t}P_{t}^{f}}
\end{equation}

Production of the non-commodity good is carried out through a one-to-one technology using domestic final goods as inputs. I assume that a perfectly competitive exporter purchases the domestic homogeneous good at price, $P_{t}$. It sells the good at dollar price, $P_{t}^{x}$, which translates into domestic currency units, $S_{t}P_{t}^{x}$. Note that competition implies that price, $S_{t}P_{t}^{x}$, equals marginal cost, $P_{t}$.

\noindent
\textbf{Commodity Exports.} The commodity side of exports is governed by stochastic processes. Both prices and quantities produced of the commodity good are determined by exogenous processes. For simplicity, I assume that the endowment process is completely constant at a steady state value of $\bar{Y}^{Co}$. I assume that the international commodity price follow an $AR(1)$ process in logarithms
\begin{align}
    \log \left(\frac{P^{Co}_t}{\Bar{P^{Co}}}\right) &= \rho^P \log \left(\frac{P^{Co}_{t-1}}{\Bar{P^{Co}}}\right) + \epsilon^P_t \label{eq:commodity_price}
\end{align}
where $P^{Co}_t$ and $\bar{P^{Co}}$ are the prices of the commodity good in period $t$ and in steady state expressed in foreign currency; and $\epsilon^{P}_t$ is a Gaussian innovations with standard deviations given by $\sigma^{P}_t$. 


The exported quantities of the sector are determined by the difference between the domestic production  and the local consumption of the commodity good
\begin{align*}
    X^{Co}_t &= Y^{Co}_t - C^{Co}_t \\
    X^{Co}_t &= Y^{Co}_t - \lambda^R C^{R,Co}_t - \left(1-\lambda^R\right) C^{H,Co}_t
\end{align*}
In the parametrization of the model in Section \ref{subsec:calibration}, parameters are chosen such that the economy is a net-exporter of the commodity good both in steady state and for reasonable values of the stochastic processes in the economy.\footnote{In other words, parameters are chosen such that in both the non-stochastic steady state of the economy and under simulations the trade balance of the commodity good is positive.}

\subsection{Policy} \label{subsec:model_policy}

In this economy government is comprised of two distinct policy makers. First, there is a monetary authority which sets a policy rule for the nominal interest rate. Second, there is a fiscal institution which carries out government expenditure of the domestic homogeneous final good. 

The central bank sets the nominal interest rate every period following a standard Taylor rule of the form
\begin{align} \label{eq:TaylorRuleBenchmark}
    \log \left(\frac{R_{d,t}}{\bar{R}_d}\right) &= \rho_R \log  \left(\frac{R_{d,t-1}}{\bar{R}_d}\right) \nonumber \\
    & \quad + \left(1-\rho_R\right) \left[ \phi_{\pi} \log\left(\frac{\lambda^R\pi^{C,R}_{t} + \left(1-\right)\pi^{C,H}_t}{\bar{\pi}} \right) + \phi_y \log\left(\frac{y_{t}}{\bar{y}} \right) + \phi_s \log \left(\tilde{S}_t\right) \right] + \epsilon^R_t
\end{align}
where $R_{d,t}$ is the central bank's risk-less interest rate, $\pi^{C,R}_t$ and $\pi^{C,H}_t$ are the Ricardian and Hand-to-mouth consumption basket inflation rates, $\bar{\pi}$ is the inflation target, $y_t = Y_t / A_t$ is final output scaled by technology, $\tilde{S}_t$ is the nominal exchange rate. and $\epsilon^R_t$ is an i.i.d. mean zero policy shock. I consider the Taylor rule described by Equation \ref{eq:TaylorRuleBenchmark} as a benchmark and study alternatives in Section \ref{sec:optimal_policy} where I study the design of optimal policies.

The fiscal authority follows a government expenditure rule every period. This rule is composed of an endogenous and exogenous block.
\begin{equation} \label{eq:fiscal_rule}
    G_t = \bar{g} + \tau^C \left(\frac{P^{Co}_t Y^{Co}_t}{\bar{P^{Co}}\bar{Y}^{Co}} -1 \right) 
\end{equation}
The first term on the RHS of Equation \ref{eq:fiscal_rule} is comprised of a constant or steady state value of government expenditure. The second term implies that government expenditure reacts to deviations of the commodity price from steady state values via a tax rate $\tau^C$.\footnote{This rule is similar to the one proposed by \cite{garcia2015dealing}. However, I abstract from government debt for simplicity.} I assume that taxes are levied lump-sum so that the resource constraint is satisfied every period. Consequently, I abstract from government debt issues.\footnote{See \cite{garcia2015dealing} on fiscal expenditure optimal policy rules which take into account government debt explicitly.} If $\tau^C<0$, then the government expenditure rule will decrease government expenditure and taxes during periods in which the total revenue of the commodity exporting sector are above their steady state values and increase it when commodity revenues fall. Given that commodity prices explain a large fraction of macroeconomic fluctuations (see \cite{drechsel2018commodity}), a rule parametrized to $\tau<0$ will take the form of a counter-cyclical government expenditure rule.

\section{Model Dynamics} \label{sec:dynamics}

This section of the paper carries out impulse response function exercises to analyze the aggregate implications of the different features of the model introduced in Section \ref{sec:model}.\footnote{The computation and solution to the model is carried out using the Dynare toolbox version 4.6.3. For more details on this toolbox see \cite{adjemian2011dynare}. Given that in Section \ref{sec:optimal_policy} I study the welfare benefits of different policy regimes, I solve the model using a second order approximation around the deterministic steady state. For more details on the solution and computation of the steady state see Appendix \ref{subsec:appendix_model_steady_state}.} In particularly, I focus on three structural shocks: a foreign interest rate shock, a commodity price shock, and a domestic monetary policy shock. Section \ref{subsec:calibration} describes the parametrization of the model based on the micro level stylized facts presented in Section \ref{sec:data} and aggregate moments. Section \ref{subsec:role_non_homotheticity} studies how introducing non-homothetic preference over the consumption of a commodity good matter for the transmission of both foreign interest rate and commodity price shocks. Section \ref{subsec:redistribution_channel} studies the aggregate and household impact of exogenous domestic monetary policy shocks households. Furthermore, I argue that under the framework presented in Section \ref{sec:model} monetary policy affects aggregates through a re-distributive channel. 

\subsection{Model Parametrization} \label{subsec:calibration}

In this section I describe the parametrization of the model. In particular, I focus on discussing the choice of households' and aggregate parameter values. This parametrization is the benchmark for the impulse response functions presented in Sections \ref{subsec:role_non_homotheticity} and \ref{subsec:redistribution_channel}. Appendix \ref{sec:appendix_calibration} presents a full description of parameter values. 

I begin by describing the parametrization of the Ricardian and Hand-to-mouth household's parameters. Overall, the rationale behind the choice of households' parameter is to simultaneously match statistics presented in Section \ref{sec:data}: the share of Ricardian and Hand-to-mouth households, and heterogeneity in the share of consumption of commodity based nourishment goods without imposing unrealistic steady state values on models' variables. Table \ref{tab:micro_calibration} presents the benchmark numerical values and description of households' key parameters.
\begin{table}[ht]
    \centering
    \caption{Benchmark Parametrization of Household Parameters}
    \label{tab:micro_calibration}
    \begin{tabular}{c|c|c}
       Parameter      & Description     & Value \\ \hline \hline 
       $\beta$        & Discount Factor & $1.05^{-1/4}$ \\
       $\lambda^R$    & Share of Ricardian HH. & 0.50 \\
       $\varphi$      & Frisch Elasticity     & 1 \\
       $\chi_R$       & Ricardian Labor Dis-utility & 1 \\
       $\chi_H$       & Hand-to-Mouth Labor Dis-utility & 1 \\
       $\alpha_{Co}$  & Cons. Share on Commodity &  0.25 \\
       $\phi_{Co}$    & Commodity Non-Homotheticity & 0.45 \\
    \end{tabular}
\end{table}
The choice of the discount factor $\beta$ is consistent with a steady state interest rate of 5\%, which is in line with EMBI spreads for Uruguay in the last two decades.\footnote{Furthermore, this value of $\beta$ is close to other values used in the literature, for instance higher than the value chosen in \cite{garcia2012real}, but lower than the value used in \cite{cubas2012rate}. The choice of a logarithmic utility function over consumption is in line with \cite{cubas2012rate} and \cite{camaraetal2021}.} The share of Ricardian households, $\lambda^R$, is set to 0.50. This motivation behind this value is twofold. Figure \ref{fig:Fact1} in Section \ref{sec:data} the first bar graph in  shows that around 50\% of Uruguayan households have access to a bank account. However, the amount of savings and/or limits to overdrafts on these banks accounts is not clear. Thus, assuming 50\% of Ricardian households is an upper bound on the share of agents with access to saving technologies. Additionally, 50\% of Ricardian households is in line with other papers in the literature of TANK models in small open economies (see \cite{garcia2015dealing} and \cite{cugat2019emerging}). I set the Frisch elasticity equal to 1, in line with \cite{mendoza2002credit}. The only parameter values which differ across households are $\chi_R$ and $\chi_H$ which are set to match steady state labor across the two type of households. Finally, I set $\alpha_{Co}$ equal to 0.25 and $\phi_{Co}$ equal to 0.45 such that in steady state commodity expenditure represent 33.69\% and 48.43\% of Ricardian and Hand-to-Mouth consumption, respectively (in line with the results presented in Table \ref{tab:Fact2}). 

Second, I describe the numerical values key model parameters that govern aggregate trade and policy rules, presented in Table \ref{tab:macro_calibration}. The first four parameters relate to the export of commodity and non-commodity goods
\begin{table}[ht]
    \centering
    \caption{Parametrization of Macro Parameters}
    \label{tab:macro_calibration}
    \begin{tabular}{c|c|c}
       Parameter      & Description     & Value \\ \hline \hline 
       $\bar{Y}^f$    & SS. Foreign Demand Shifter & 5 \\
       $\eta^f$       & Non-Commodity Export Elasticity & 1.5 \\       
       $\bar{Y}^{Co}$ & SS. Commodity Endowment    & 2 \\
       $\bar{P}^{Co}$ & SS. Commodity Price    & 0.15 \\
       $\bar{\pi}$    & Central Bank's Inflation Target & $1.05^{1/4}$ \\
       $\rho_R$       & Auto-Regressive Parameter Taylor Rule & 0.75 \\
       $\phi_{\pi}$   & Inflation Coefficient Taylor Rule & 1.5 \\
       $\phi_{y}$     & GDP Gap Coefficient Taylor Rule & 0.05 \\
       $\phi_{s}$     & Exchange Rate Coefficient Taylor Rule & 0.02 \\
    \end{tabular}
\end{table}
This parameter choice implies a steady state ratio of total exports to GDP of 33\%, with 80\% of exports being commodity goods in line with aggregate trade data. Additionally, the steady state price of the commodity good is set to 0.15, in line with firm-customs level data on export prices of commodity and non-commodity goods.\footnote{For more details on aggregate and firm-customs level data see Appendix \ref{subsec:appendix_firm_findings} for more details.}  The bottom five parameters of Table \ref{tab:macro_calibration} govern the central bank's Taylor rule. The inflation target is set to 5\%, in the middle of the 3\% to 7\% inflation rate set by the Uruguayan Central Bank.\footnote{See \url{https://www.bcu.gub.uy/Politica-Economica-y-Mercados/Paginas/Reportes-de-Politica-Monetaria.aspx}. This value is in line with the value used in \cite{cubas2012rate}.} The numerical values of the Taylor rule's reaction coefficients $\rho_R, \phi_{\pi}, \phi_{y}, \phi_{s}$ are in line with the estimation results of \cite{christiano2011introducing} and \cite{gonzalez2014towards}. In terms of fiscal policies, the value of parameter $\bar{g}$ is set to match a government expenditure to GDP ratio of 30\% in line with data coming from the IMF's WEO. For the benchmark exercises carried out in Sections \ref{subsec:role_non_homotheticity} and \ref{subsec:redistribution_channel} $\tau^{C}$ is set to zero.

\subsection{Non-Homotheticity \& Foreign Shocks} \label{subsec:role_non_homotheticity}

Next, I describe the role of non-homothetic preferences over commodity goods in propagating two relevant exogenous shocks in small open economies: a foreign interest rate shock $R^{*}_t$ and a commodity price shock $P^{C}_t$. These exercises shed light on how changes in prices of key inputs of households' consumption baskets have aggregate repercussions. In particular, I argue that higher degrees of non-homotheticity amplifies the macroeconomic impact of these shocks.

\begin{figure}[ht]
    \centering
    \caption{ Foreign Interest Rate Shock $R^{*}_{t}$}
    \label{fig:aggregate_Rstard_P}
    \includegraphics[width=16cm, height=12cm]{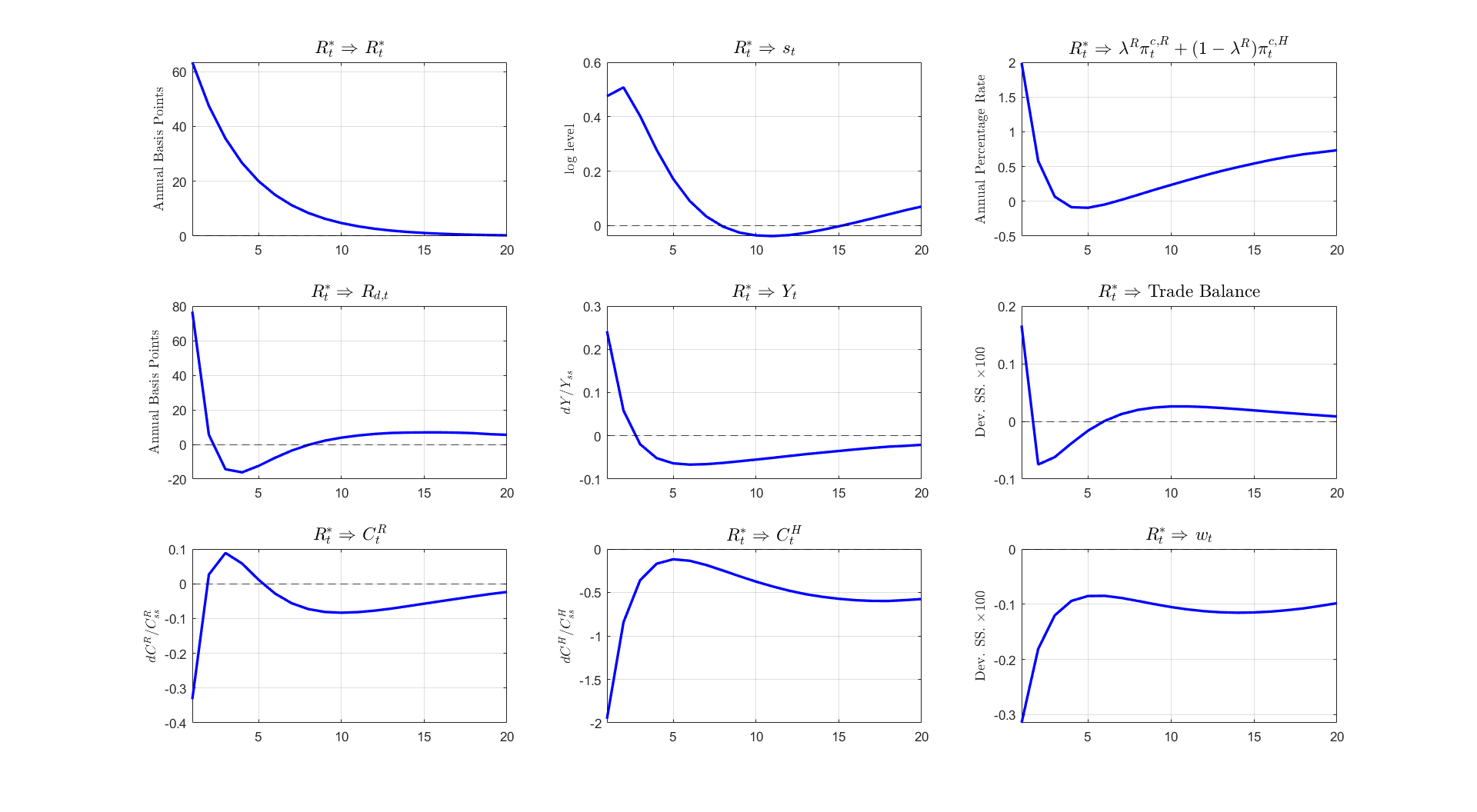}
\end{figure}

\noindent
\textbf{Foreign Interest Rate Shock.} First, I study the impulse response functions of a foreign interest rate shock on this small open economy. Figure \ref{fig:aggregate_Rstard_P} presents the impulse response functions dynamics.
The rise in the foreign interest rate first impacts the small open economy through the uncovered interest rate parity condition. An increase of 60 annual basis points (see Panel 1.1) leads to a depreciation of the nominal exchange rate (see Panel 1.2). The exchange rate depreciation increases in the domestic currency price of both commodity and imported consumption goods, leading to a jump in the consumption bundle inflation rate of both households (see Panel 1.3). The central bank's policy rate reacts to the exchange rate pass-through in prices by increasing the nominal policy rate $R_{d,t}$ (see Panel 2.1). While the initial depreciation boost exports and the production of the domestic final good $Y_t$ for the first two periods after the shock, the higher domestic rate leads to a persistent drop in $Y_t$ even 20 periods after the shock. The trade balance exhibits a similar pattern, with a sharp improvement in the first period which rapidly becomes a deficit, and later returns to its steady state value (see Panel 2.3). Ricardian and Hand-to-mouth households' consumption exhibit a significant drop on impact (see Panels 3.1 and 3.2 respectively). However, Hand-to-mouth's consumption a significantly greater initial drop and slower recovery. This slow recovery is primarily explained by the persistent drop in real wages measured in terms of the domestic final good (see Panel 3.3).

\begin{figure}[ht]
    \centering
    \caption{ Foreign Interest Rate Shock $R^{*}_{t}$ \\ Role of Non-Homothetic Preferences}
    \label{fig:HH_Rstard_P}
    \includegraphics[width=16cm, height=10cm]{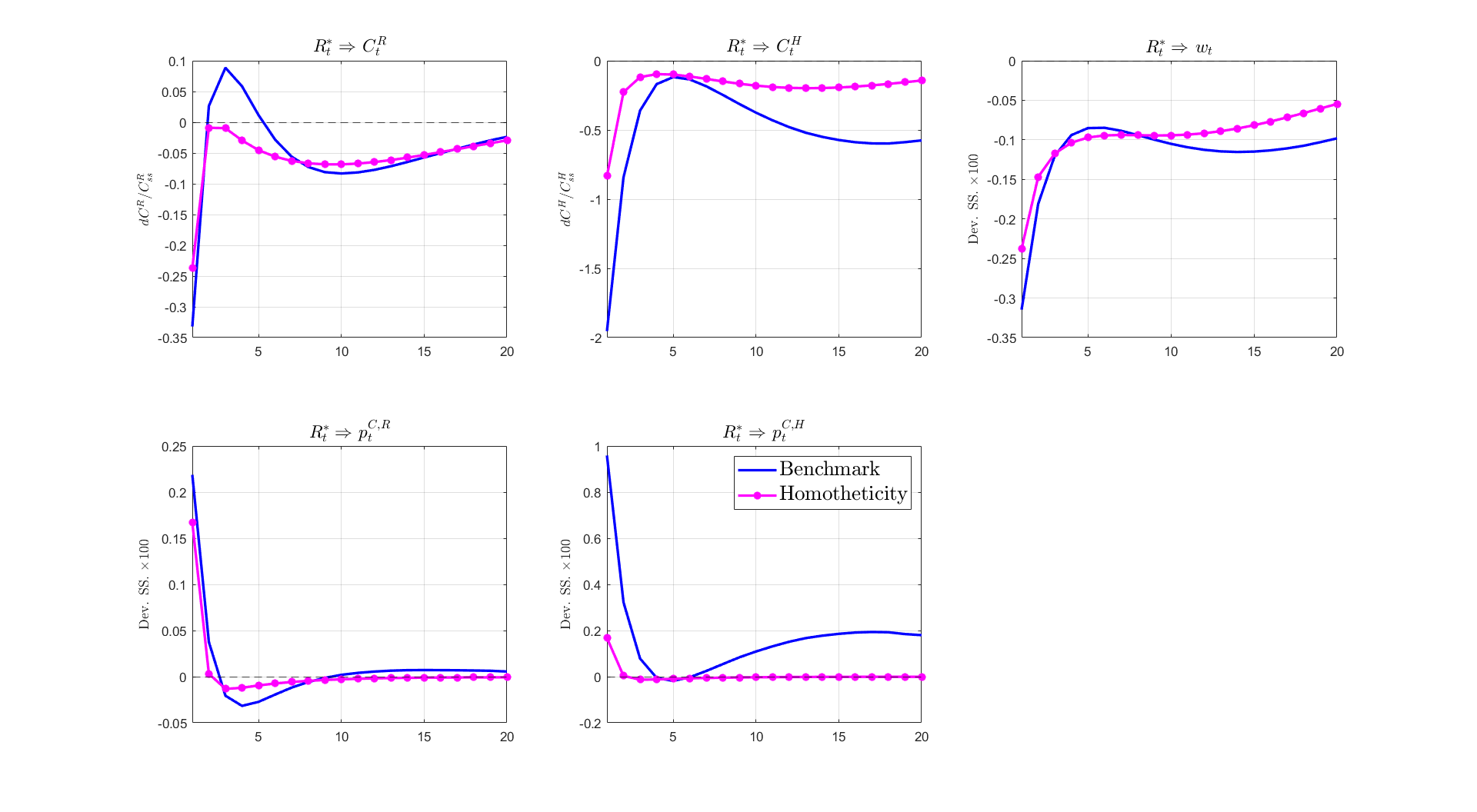}
\end{figure}

Next, I study the implications of non-homothetic preferences in the response of household specific variables to a foreign interest rate shocks $R^{*}_t$. Figure \ref{fig:HH_Rstard_P} shows the dynamics of household specific consumption, consumption price bundles and real wages for the benchmark model (blue solid line) and a parametrization homothetic preferences (magenta dotted line).
Non-homothetic preferences barely affect response of Ricardian consumption to a foreign interest rate shock $R^{*}_t$ (see Panel 2.1). However, non-homothetic preferences lead to a significantly greater and more persistent drop in Hand-to-Mouth consumption (see Panel 2.2). These greater initial drop in Hand-to-Mouth consumption in the model with non-homothetic preferences is partially explained by a greater drop in real wages, measured in terms of domestic final good. Notwithstanding, non-homothetic preferences impact households' consumption primarily through price indexes. Panel 2.1 of Figure \ref{fig:HH_Rstard_P} shows that the relative price of the Ricardian consumption bundle with respect to the domestic final good hardly changes with respect to a homothetic exercise. However, Panel 2.2 shows that the increase in the relative price of Hand-to-mouth consumption bundle is five times greater with non-homothetic preferences compared to the homothetic exercise.

In summary, the model's benchmark parametrization leads to aggregate dynamics in line with other Two Agent New Keynesian models. The introduction of non-homothetic preferences amplifies the dynamics of both households' consumption. The Hand-to-mouth household is particularly hit due to the drop in real wages and the lack of access to saving technologies. Moreover, non-homothetic preferences implies that as consumption decreases, a greater share of income is spent on the now more expensive commodity good (due to exchange rate pass-through). Thus, non-homothetic preferences amplify the drop in Hand-to-Mouth consumption. 

\begin{figure}[ht]
    \centering
    \caption{ Commodity Price Shock $P^{Co}_{t}$}
    \label{fig:aggregate_Pco}
    \includegraphics[width=16cm, height=12cm]{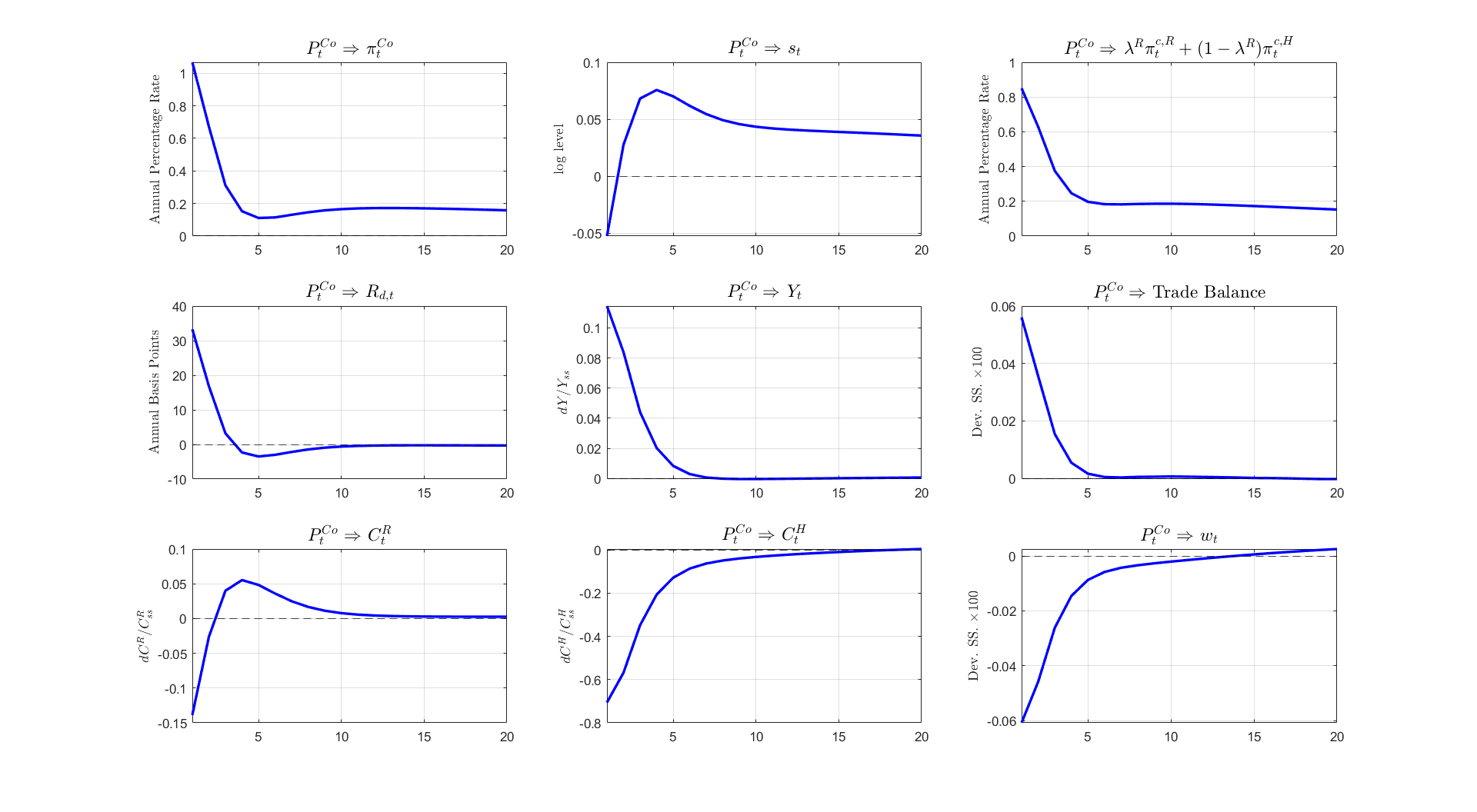}
\end{figure}

\noindent
\textbf{Commodity Price Shocks.} Next, I turn to analyzing the model's dynamics after a commodity price shock. Figure \ref{fig:aggregate_Pco} presents the impulse response functions dynamics.
The higher international prices for the commodity good impact the small open economy through two main channels: a transitory wealth effect and an increase in the price of households' consumption bundle. A 5\% international commodity price shock leads to a 1\% increase in the domestic currency price of the commodity good (see Panel 1.1). The smaller impact on the domestic currency price of the commodity good is explained by the initial sharp appreciation of the nominal exchange rate (see Panel 2.1). However, the nominal appreciation is short-lived, leading to a mild depreciation in the following periods. The rise in the commodity price leads to an increase in the consumption bundle inflation rate (see Panel 1.3). The monetary authority reacts by increasing the nominal interest rate $R_{d,t}$ (see Panel 2.1). On the real side of the economy, the production of the domestic final good expands and the trade balance experiences a surplus (see Panels 2.2 and 2.3). On the one hand, Ricardian consumption experiences a initial drop and a mild expansion in subsequent periods, explained by consumption smoothing to a transitory wealth shock. On the other hand, the commodity price shock leads to a significant drop in Hand-to-Mouth consumption. Finally, Panel 3.3 shows that drop in Hand-to-Mouth consumption is at least partially explained by a drop in real wages measured in terms of the domestic final good. 

Again, I analysis the implications of non-homothetic preferences in the response of household specific variables to a commodity price shock $P^{Co}_t$.
\begin{figure}[ht]
    \centering
    \caption{ Commodity Price Shock $P^{Co}_{t}$ \\ Role of Non-Homothetic Preferences}
    \label{fig:HH_Pco}
    \includegraphics[width=16cm, height=10cm]{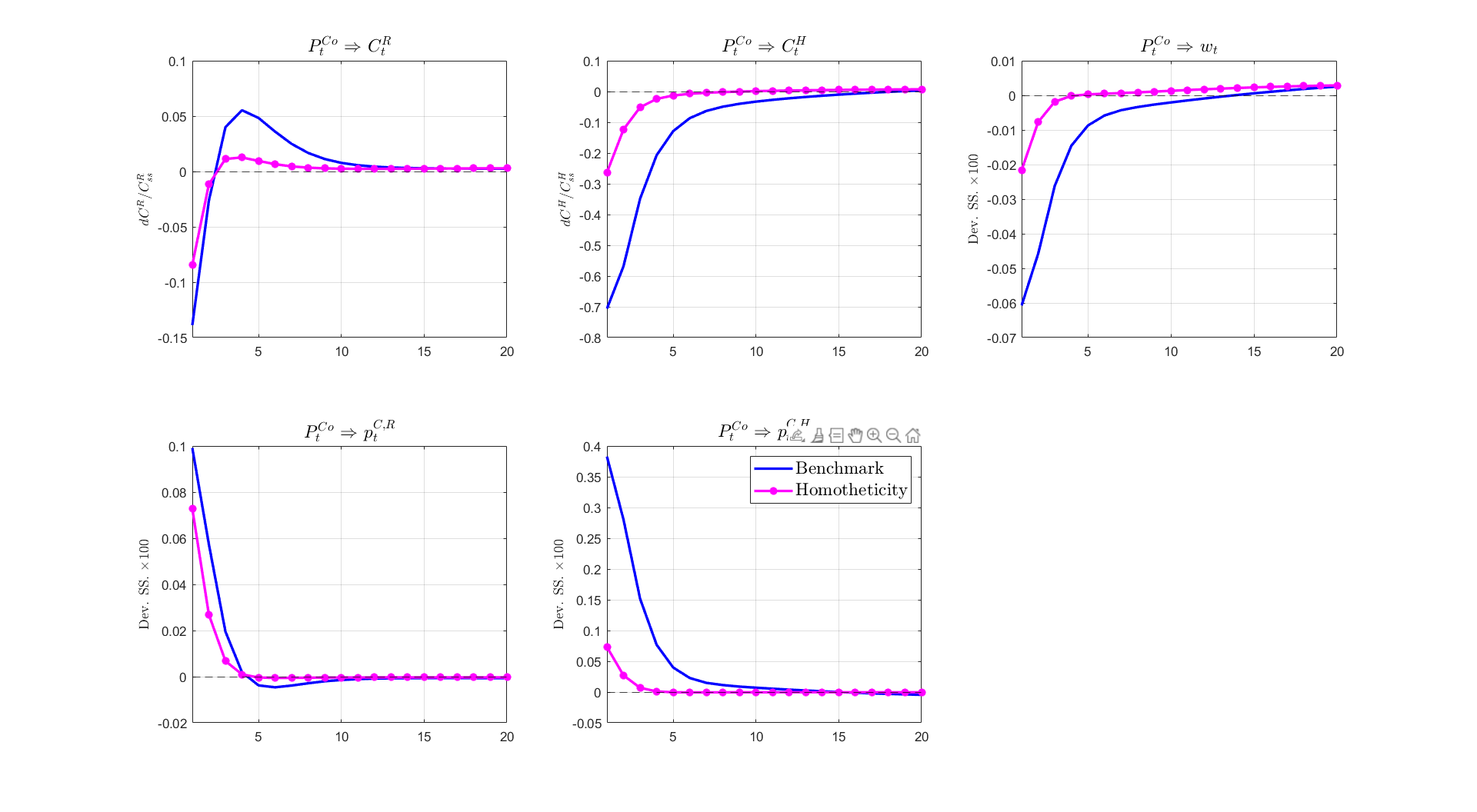}
\end{figure}
Figure \ref{fig:HH_Pco} presents the dynamics for the benchmark parametrization (blue solid line) and for a homotheticity exercise (magenta dotted line). In line with the results presented for a foreign interest rate shock, non-homothetic preferences primarily affect households which do not have access to saving technologies. To see this, note that under the benchmark parametrization, Ricardian consumption experience a sharper drop on impact, but recovers faster than the case with homothetic preferences (see Panel 1.1). But non-homothetic preferences lead to a drop in Hand-to-Mouth consumption which is almost three times greater than in the homothetic exercise. The drop in Hand-to-Mouth consumption is driven by a significant fall in real wages, measured in domestic final goods, and a greater increase in the price of its consumption bundle.

In short, while the benchmark model parametrization leads to aggregate dynamics in line with a previous literature, non-homothetic preferences matter for household specific dynamics. While Ricardian households experience an increase in consumption after an increase in commodity prices, Hand-to-Mouth households experience a persistent drop in consumption. Moreover, I show that non-homothetic preferences amplify the negative impact of a commodity price shock on Hand-to-Mouth households. These results are in line with a literature which has highlighted how the commodity price boom in the 2000s negatively affected poor income households in Latin American (see \cite{frenkel2016inflacion} for Argentina and \cite{estrades2012commodity} for Uruguay) and African (\cite{benson2008impacts} for Uganda and \cite{shimeles2013rising} for Ethiopia) countries. Thus, this framework highlights an underlying tension which arises in agricultural commodity exporting countries: an increase in commodity prices creates a positive wealth shock for the small open economy but negatively affects poor income countries which spend a significant share of their income in food and have no access to saving technologies.

\subsection{Redistribution Channel of Domestic Monetary Policy} \label{subsec:redistribution_channel}

In this section of the paper I show that in the structural framework presented in Section \ref{sec:model} monetary policy has a redistribution channel. More specifically, exogenous shocks in the domestic monetary policy rate produce opposite dynamics for the consumption of Ricardian and Hand-to-mouth households. Furthermore, I argue that the relative importance of this redistribution channel increases with the degree of non-homotheticity over the consumption of commodity goods. In order to show this, I carry out impulse response function exercises of a domestic monetary policy shock under different model parametrizations.

\begin{figure}[ht]
    \centering
    \caption{ Domestic Monetary Policy Shock $\epsilon^{R}_{t}$}
    \label{fig:aggregate_R}
    \includegraphics[width=16cm, height=12cm]{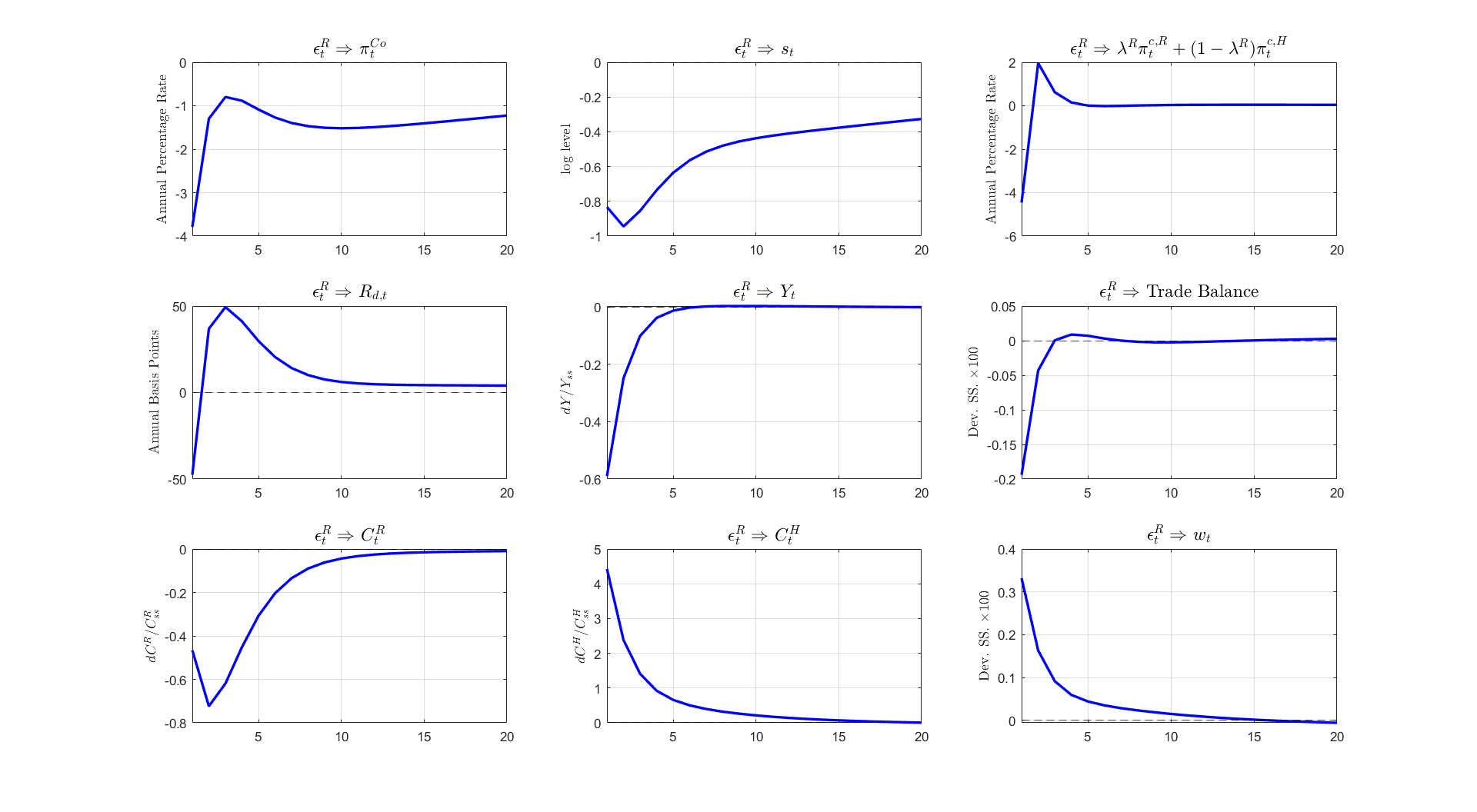}
\end{figure}

First, I study the dynamics of benchmark model after a domestic interest rate shock, presented in Figure \ref{fig:aggregate_R}.
This shock leads to a persistent drop in the domestic currency commodity price (see Panel 1.1) explained by a persistent nominal exchange rate appreciation (see Panel 1.2). The drop in the domestic currency price of the commodity good leads to an overall initial fall in the consumption price bundle inflation (see Panel 1.3). Panel 2.1 exhibit the dynamics of the domestic monetary policy interest rate $R_{d,t}$. The significant drop in the domestic price of the commodity good leads to an initial and brief drop in $R_{d,t}$ which later reverts to a temporary increase. The higher interest rate lead to a drop in the production of the homogeneous final good (see Panel 2.2) and a deterioration of the trade balance (see Panel 2.3). Households' consumption experience exactly opposite dynamics. On the one hand, the Ricardian household experience a hump-shaped decrease in consumption driven by inter-temporal consumption smoothing (see Panel 3.1). On the other hand, Hand-to-mouth households experience a significant increase in consumption (see Panel 3.2). This increase in $C^{H}_t$ is explained by both the decrease in the domestic price of commodity goods and an increase in real wages measured in terms of the homogeneous final good (see Panel 3.3). In summary, the structural framework presents aggregate dynamics in line with previous literature (exchange rate appreciation, fall in inflation, drop in economic output), but suggests households react differently to a hike in domestic interest rates.

Next, I study the role of non-homothetic preferences in explaining the opposing households' consumption dynamics in response to a domestic monetary policy shock. 
\begin{figure}[ht]
    \centering
    \caption{ Domestic Monetary Policy Shock $\epsilon^{R}_{t}$ \\ Role of Non-Homothetic Preferences}
    \label{fig:HH_R}
    \includegraphics[width=16cm, height=10cm]{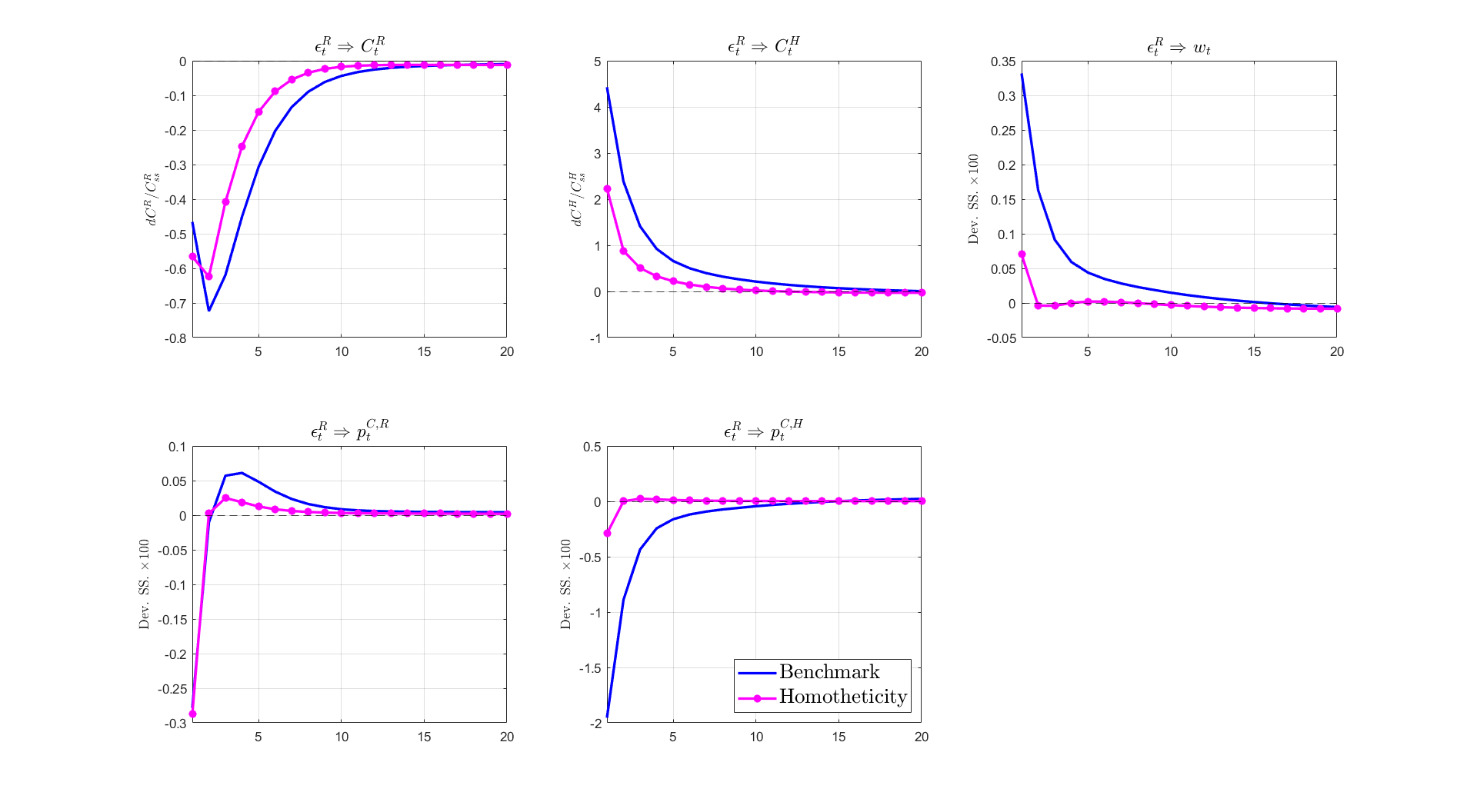}
\end{figure}
Figure \ref{fig:HH_R} presents the dynamics for the benchmark parametrization (blue solid line) and for a homotheticity exercise (magenta dotted line). Note that non-homothetic preferences are not the driving force which explains the opposing consumption dynamics. The inter-temporal smoothing channel leads Ricardian households to reduce consumption under both exercises (see Panel 1.1). Hand-to-mouth households increase consumption (see Panel 1.2) due to the increase in real wages (see Panel 1.3) and the reduction in the consumption price bundle (see Panel 2.2). However, in line with the results presented in Section \ref{subsec:role_non_homotheticity}, non-homothetic preferences matter for the quantitative impact of a domestic monetary policy shock for Hand-to-mouth households. Panel 1.2 shows that the presence of non-homothetic preferences doubles the increase of Hand-to-mouth consumption, while a significantly smaller impact for Ricardian consumption (see Panel 1.1). The greater increase in $C^{H}_t$ is driven by an appreciably greater increase in real wages and significantly sharper decrease in the Hand-to-mouth consumption price index.\footnote{Note that non-homothetic preferences do not affect the Ricardian consumption price index as significantly as the Hand-to-mouth consumption price index}

To summarize, the exercises presented in this section showed that under the present structural framework a domestic monetary policy shock impact households' consumption with opposite signs. Hand-to-mouth households are benefited by the nominal exchange rate appreciation which reduces the domestic currency price of the consumption commodity good. Additionally, I showed that non-homothetic preferences matter for the quantitative impact of a domestic monetary policy shock on Hand-to-mouth households' variables. These results suggest that monetary policy has a re-distributional channel, negatively affecting Ricardian households while benefiting Hand-to-mouth households. 

\section{Optimal Policy} \label{sec:optimal_policy}

This section of the paper computes optimal monetary and fiscal policies from the point of view of both Ricardian and Hand-to-mouth households. I begin by analyzing how different policy regimes affect both aggregate and household specific dynamics in response to exogenous foreign shocks. Then, I compute household specific preferences over the different policy regimes using a second order approximation to the welfare function as in \cite{schmitt2004solving}. I show that Ricardian and Hand-to-mouth households have opposing preferences over policy regimes. Finally, I shed light on the reasoning behind these differences in preferred policy regimes by analyzing how they shape the volatility of households' consumption and labor profiles and stress the role of non-homothetic preferences over commodity goods. 

\subsection{Dynamics under Different Policy Rules} \label{subsec:policy_rules}

Before carrying out the welfare analysis I show how different policy regimes shape aggregate and household specific dynamics in response to foreign interest rate shocks and commodity price shocks. This exercise sheds light on how households value different policy regimes. I consider both monetary and fiscal policy regimes.

\begin{figure}[ht]
    \centering
    \caption{  Foreign Interest Rate Shock $R^{*}_{t}$ \\ Different Monetary Regimes}
    \label{fig:rS}
    \includegraphics[width=16cm, height=12cm]{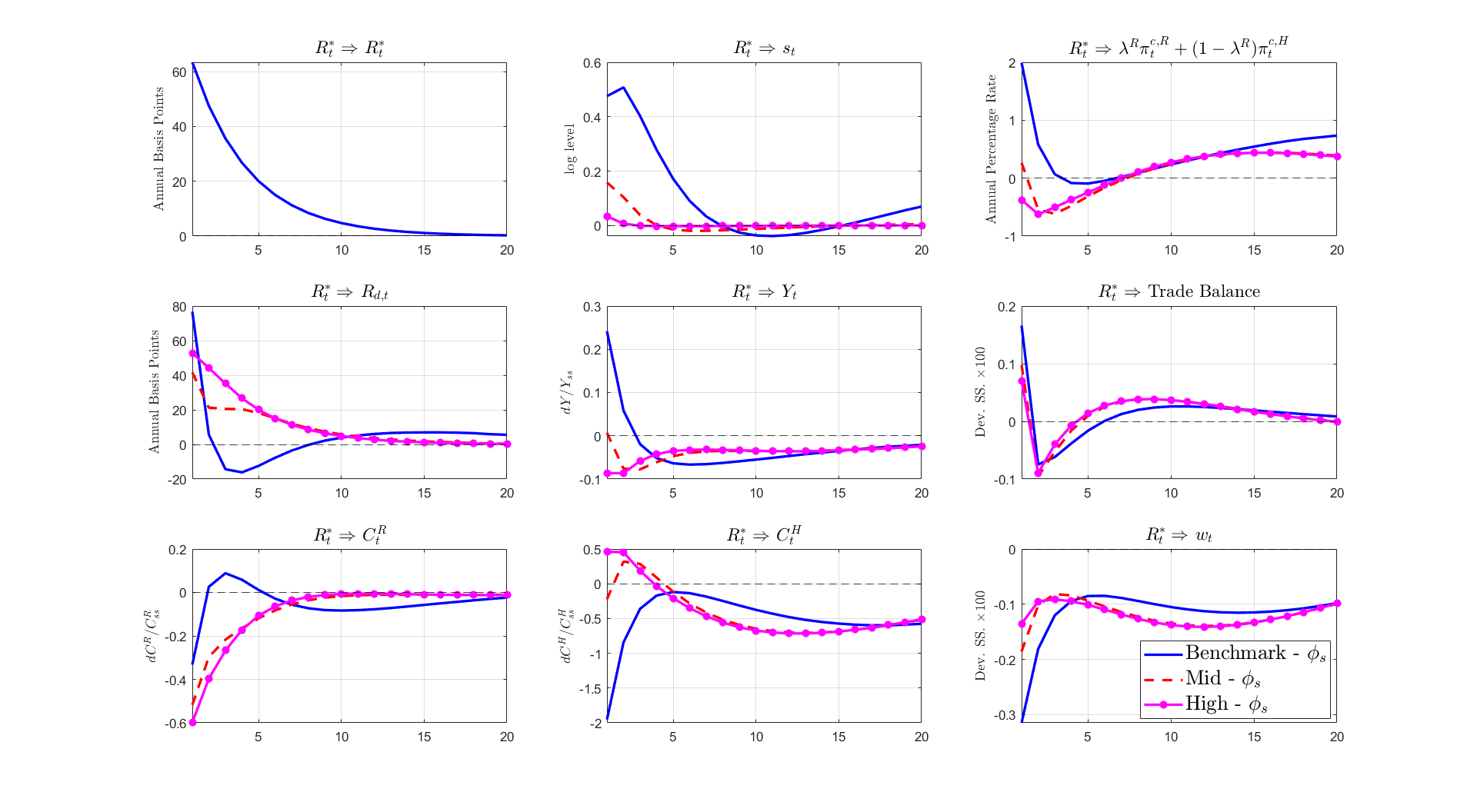}
\end{figure}

\noindent
\textbf{Flexible vs fixed exchange rate regimes.} First, I consider different monetary policy regime which differ on their exchange rate flexibility. In particular, I compare the dynamics under the benchmark parametrization presented in Figure \ref{fig:aggregate_Rstard_P} with a monetary policy regime in which the domestic rate, $R_{d,t}$, increases to reduce changes in the nominal exchange rate. Figure \ref{fig:rS} shows this comparison of the benchmark dynamics (solid blue line), a regime with a mid (red dashed line) and high value (dotted magenta line) $\phi_s$.

The impact of monetary police regimes with less-flexible exchange rate regimes are in line with a previous literature. In response to a foreign interest rate shock, regimes with a less flexible nominal exchange rate (see Panel 1.2), the impact on consumer price inflation is lower or may even lead to deflation (see Panel 1.3). The policy rate $R_{d,t}$ shows a more persistent increase (see Panel 2.1). Note that the impact on the real interest rate (proxied by the difference between the dynamics shown in Panels 2.1 and 1.3) is appreciably greater in monetary policy regimes with less flexible nominal exchange rates. The higher domestic interest rate leads to greater drops in the production of the domestic final good (see Panel 2.2) and smaller trade balance surplus (see Panel 2.3).

Following the description of aggregate dynamics, I analyze household specific dynamics under a monetary policy regime with less flexible exchange rates. On the one hand, panel 3.1 shows that, compared to the benchmark parametrization, regimes with less flexible nominal exchange lead to a greater drop in Ricardian households' consumption. This is driven by the greater response of the policy interest rate $R_{d,t}$ which leads to a higher inter-temporal motive. On the other hand, Panel 3.2 shows that Hand-to-mouth households' consumption actually increases on impact under less flexible regimes. This initial increase is driven by a drop in the consumption price index (see Panel 1.3) and lower drop in real wages measured in the domestic final good (See Panel 3.3). However, this increase is short lived, and after 4 quarters consumption decreases and follows dynamics similar to the benchmark exercise. The opposing impact over households' consumption dynamics under less flexible exchange rate regimes is in line with the results presented in Section \ref{subsec:redistribution_channel}. The benchmark exercise with a greater nominal depreciation and lower real interest rate leads to a lower impact on Ricardian households' consumption and in the production of final goods. However, the less flexible regimes reduces the impact on Hand-to-mouth's consumption price index which may even increase their total income (at least initially).

\noindent
\textbf{Pro-cyclical vs counter-cyclical regimes.} I now turn to analyzing how fiscal policy can shape aggregate and household specific dynamics in response to commodity price shocks. 
\begin{figure}[ht]
    \centering
    \caption{ Commodity Price Shock $P^{Co}_{t}$ \\ Different Fiscal Regimes}
    \label{fig:tau_co}
    \includegraphics[width=16cm, height=12cm]{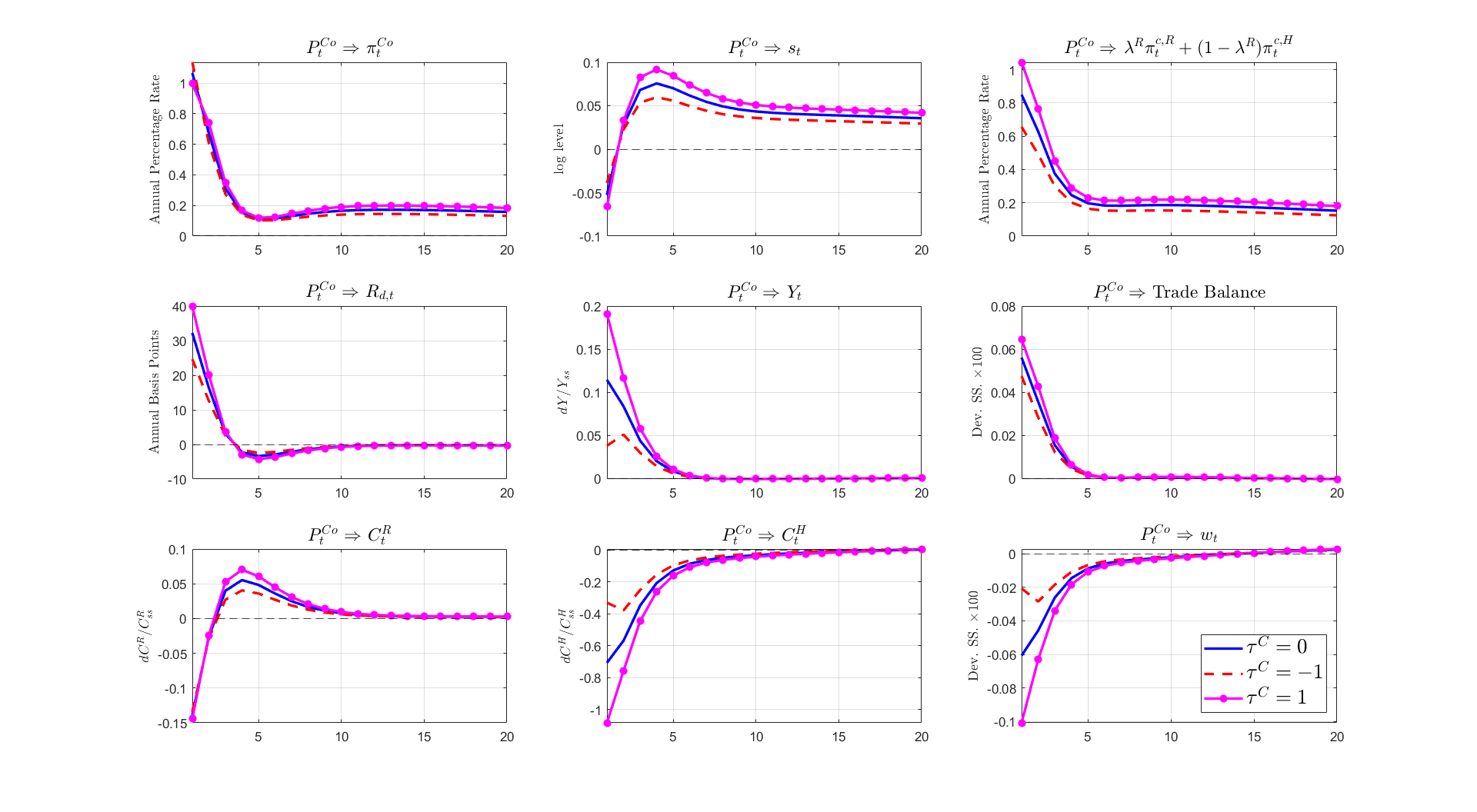}
\end{figure}
In particular, in Figure \ref{fig:tau_co} I compare the benchmark dynamics presented in Figure \ref{fig:aggregate_Pco} with two fiscal regimes depending on the value of $\tau^C$ in Equation \ref{eq:fiscal_rule}: (i) a pro-cyclical regime with $\tau^{C}=1$ (dotted magenta line) in which government expenditure increases in response to an increase in the commodity price; (ii) a counter-cyclical regime with $\tau^{C}=-1$ (dashed red line) in which government expenditure decreases in response to an increase in the commodity price. 

First, a pro-cyclical (counter cyclical) fiscal regime exacerbates (reduces) the aggregate impact of a commodity price shock, compared to the benchmark exercise. For instance, a pro-cyclical fiscal rule leads to a greater impact in consumer price in inflation (see Panel 1.3), and consequently a greater reaction of the domestic monetary interest rate (see Panel 2.1). Furthermore, the pro-cyclical rule increases aggregate demand leading to higher production of domestic final goods (see Panel 2.2). These exacerbation of aggregate dynamics can also be observed in the a greater trade balance surplus and a sharper drop in real wages measured in terms of the domestic final good (see Panels 2.3 and 3.3 respectively).

Second, I show that the pro-cyclical (counter-cyclical) fiscal regime also exacerbates (reduces) the impact of a commodity price shock on households' consumption dynamics. Panel 3.1 shows that the pro-cyclical regime leads to greater Ricardian consumption, while Panel 3.2  shows that this regime leads to a greater drop in Hand-to-mouth consumption. This greater drop is explained by the aforementioned greater consumer price inflation and the sharper drop in real wages.

In summary, a pro-cyclical fiscal policy regimes in response to commodity prices shocks amplifies the dynamics found under the benchmark parametrization presented in Section. A counter-cyclical fiscal policy rule has the opposite effect, reducing the impact of a commodity price shocks. This result implies significant differences for the households in this economy. On the one hand, Ricardian households experience an even greater increase in consumption under a pro-cyclical rule compared to the benchmark scenario. On the other hand, Hand-to-mouth households experience an even greater drop in consumption under a pro-cyclical expenditure rule compared to the benchmark scenario. Consequently, households' consumption dynamics differ across fiscal policy rules, as is the case for monetary policy regimes.

\subsection{Households' Preferences over Policy Regimes} \label{subsec:hh_policy_regimes}

This section of the paper presents a normative analysis of households' welfare under the different policy regimes described in previous sections. To do so, I choose the value of $\eta = \{\phi_s, \tau^C\}$ which maximizes 
\begin{align*}
    \mathcal{U}^j = \mathbb{E}_0 \sum^{\infty}_{t=0} \beta^{t}\left[ \log \left(C^j_t \left(\eta \right) \right) - \chi_j \frac{\left(l^j_t \left(\eta \right)\right)^{1+\varphi^j}}{1+\varphi^j} \right]
\end{align*}
for $j=\{H,R\}$. I approximate the value of this expected utility using a second order Taylor approximation around the non-stochastic steady state, following \cite{schmitt2007optimal}. Furthermore, I compute both the mean and standard deviation of households' consumption and labor across different values of $\eta$ to shed light on the household specific preferences over policy regimes.\footnote{To carry out the computation of the optimal policy exercise I impose bounds on parameter values. In particular, I set $\tau^C \in [0,1]$, which implies I consider fiscal policies anywhere between the full pro-cyclical and complete counter-cyclical cases. These bounds are in line with the optimal policy exercise carried out by \cite{garcia2015dealing}. In the case of $\phi_s$, I only set a lower bound of 0.0002. This is due to a second order approximation solution to the model fails the Blanchard-Kahn conditions for values under this lower bound. I show that both Ricardian and Hand-to-mouth's welfare is maximized at values of $\phi_s$ above this lower bound. I take this result as evidence that imposing a lower bound on $\phi_s$ does not drive the results of this normative analysis.} Finally, I carry out the same optimal policy analysis shutting down non-homothetic preferences, i.e., setting $\phi_{Co}$ equal to zero.

The results of the optimal policy regimes are presented in Table \ref{tab:welfare}. The main message of this table is that Ricardian and Hand-to-mouth households have opposing preferences over policy regimes. The former prefer regimes with strongly pro-cyclical fiscal policy and relatively more free floating exchange rate regimes, while the latter prefer counter-cyclical fiscal policy and a monetary policy regimes which exhibit ``\textit{fear-of-floating}''. The first panel shows the results for Ricardian households. For both the case with non-homothetic preferences and the case with homothetic preferences, Ricardian households prefer a pro-cyclical fiscal policy. The optimal combination of policy regimes reduce both the volatility of consumption and labor, while simultaneously increasing the mean consumption stream and decreasing average labor (compared to the benchmark case with an acyclical fiscal policy regime and a freely floating exchange rate regime).
\begin{table}[ht]
    \centering
    \caption{Optimal Policies across Households}
    \label{tab:welfare}
    \small 
    \begin{tabular}{l c c c c c c}
         & & &  \multicolumn{2}{c}{St. Dev.} & \multicolumn{2}{c}{Mean}  \\ \hline 
         & Optimal $\tau^C$ & Optimal $\phi_s$  & $C^j$ & $l^j$ & $C^j$ & $l^j$ \\ \hline \hline 
    \multicolumn{7}{c}{ } \\
    \multicolumn{7}{l}{\textbf{Ricardian Households}}  \\
    Non-Homothetic Preferences & 1 & 0.0822 & 0.8612 & 0.8159 & 0.0292 &  -0.0267 \\
    Homothetic Preferences & 1 &  1.4448 & 0.6189 & 0.3524 & 0.0386  & -0.0019 \\
    \multicolumn{7}{c}{ } \\
    \multicolumn{7}{l}{\textbf{Hand-to-mouth Households}}  \\
    Non-Homothetic Preferences & -1 & 11.7048 & 0.4074 & 0.3929 & 0.1590 & -0.1880 \\
    Homothetic Preferences & -1 & 34.7898 & 0.3415 & 0.4016 & 0.0503 & -0.0409 \\
    \multicolumn{7}{c}{ } \\ \hline \hline 
    \end{tabular}
    \floatfoot{\footnotesize \textbf{Note}: The second and third line present the  household specific values of $\tau^C$ and $\phi_s$ which maximizes ex ante utility, respectively. The fourth column, ``St. Dev. $C^j$'' presents the ratio of standard deviation of consumption under the optimal values of vector $\eta$ compared to the case of the benchmark parametrization with $\phi_s = 0.02$ and $\tau^C = 0$. The fifth column, ``St. Dev. $l^j$'' is the analogous exercise but for households' labor. The sixth column, ``Mean $C^j$'', represents the percentage increase in the mean household consumption under the optimal set $\eta$ relative to the benchmark parametrization. The last column, ``Mean $l^j$'', carries out the analogous exercise for households' labor. The computation of these exercises is carried out using a second-order approximation around the non-stochastic steady state. In the case with homothetic preferences, $\phi_{Co}$ is set to zero and $\chi_R$ and $\chi_H$ are chosen such that households work the same amount of time in steady state.}
\end{table}

\begin{figure}[ht]
    \centering
    \caption{Ricardian Consumption \& Policy Regimes \\ \footnotesize $C^R$ as a function of $\eta$}
    \label{fig:cR_mean_stdev}
    \includegraphics[width=14cm, height=6cm]{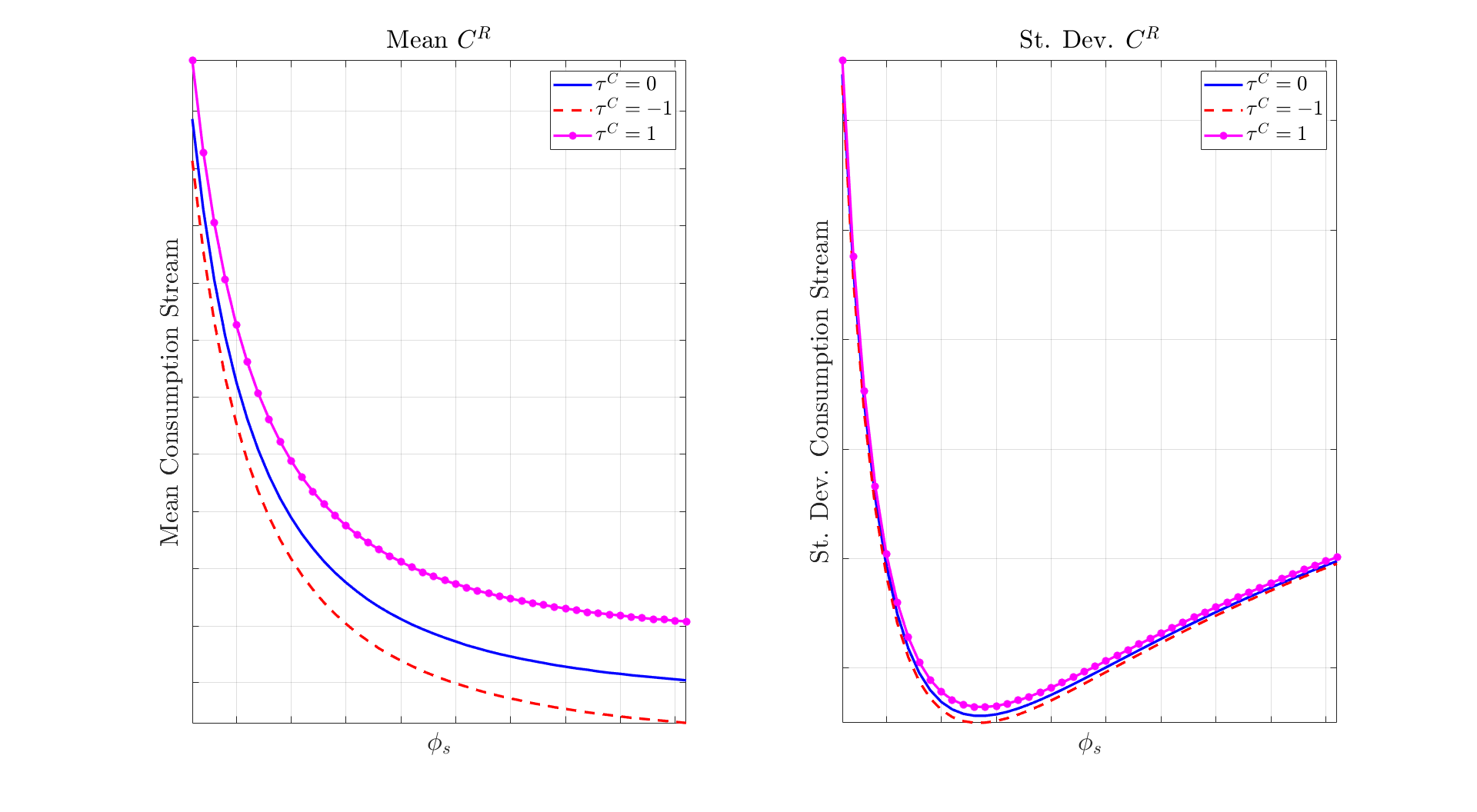}
    \floatfoot{\footnotesize \textbf{Note}: The computation exercise is carried out for the benchmark specification with non-homothetic preferences. Results are qualitatively similar for the exercise without non-homothetic preferences. }
\end{figure}

To better understand the intuition behind Ricardian households' preferences over policy regimes, Figure \ref{fig:cR_mean_stdev} shows the mean and standard deviation of Ricardian consumption streams as a function of $\eta$.
The first panel shows that pro-cyclical fiscal rules lead to higher mean Ricardian consumption streams, compared to acyclical or counter-cyclical rules. Also, for a given fiscal rule, higher exchange rate rigidity decreases mean consumption streams. The second panel shows that pro-cyclical fiscal rules increases the volatility of consumption streams compared to acyclical or counter-cyclical rules.
Interestingly, the exchange rate rigidity has a non-monotonic relationship . For a given fiscal policy rule, an increase in exchange rate rigidity initially reduces consumption volatility, but after some intermediate level, further increasing $\phi_s$ leads to greater volatility. Consequently, the Ricardian households' optimal set of policy regimes is the fiscal policy which maximizes mean consumption streams. This fiscal policy rule is accompanied of a low-degree of exchange rate rigidity which reduces overall consumption volatility, albeit sacrificing a small degree of mean consumption stream.

Next, I turn to the analysis of optimal policy regimes from the perspective of the Hand-to-mouth households, shown in the second panel of Table \ref{tab:welfare}. Contrary to Ricardian households, Hand-to-mouth households prefer counter-cyclical fiscal policies for both the non-homothetic and homothetic cases. Furthermore, Hand-to-mouth households with and without non-homothetic preferences have preference towards ``dirty-float'' exchange rate regimes. This set of optimal policy regimes lead to a significant reduction in Hand-to-mouth consumption and labor compared to the benchmark parametrization (see columns 4 and 5 respectively). Additionally, the optimal policy regime leads to a significant increase in the mean consumption stream and a decrease in mean labor compared to the benchmark parametrization.

\begin{figure}[ht]
    \centering
    \caption{Hand-to-mouth Consumption \& Policy Regimes \\ \footnotesize $C^H$ as a function of $\eta$}
    \label{fig:cH_mean_stdev}
    \includegraphics[width=14cm, height=6cm]{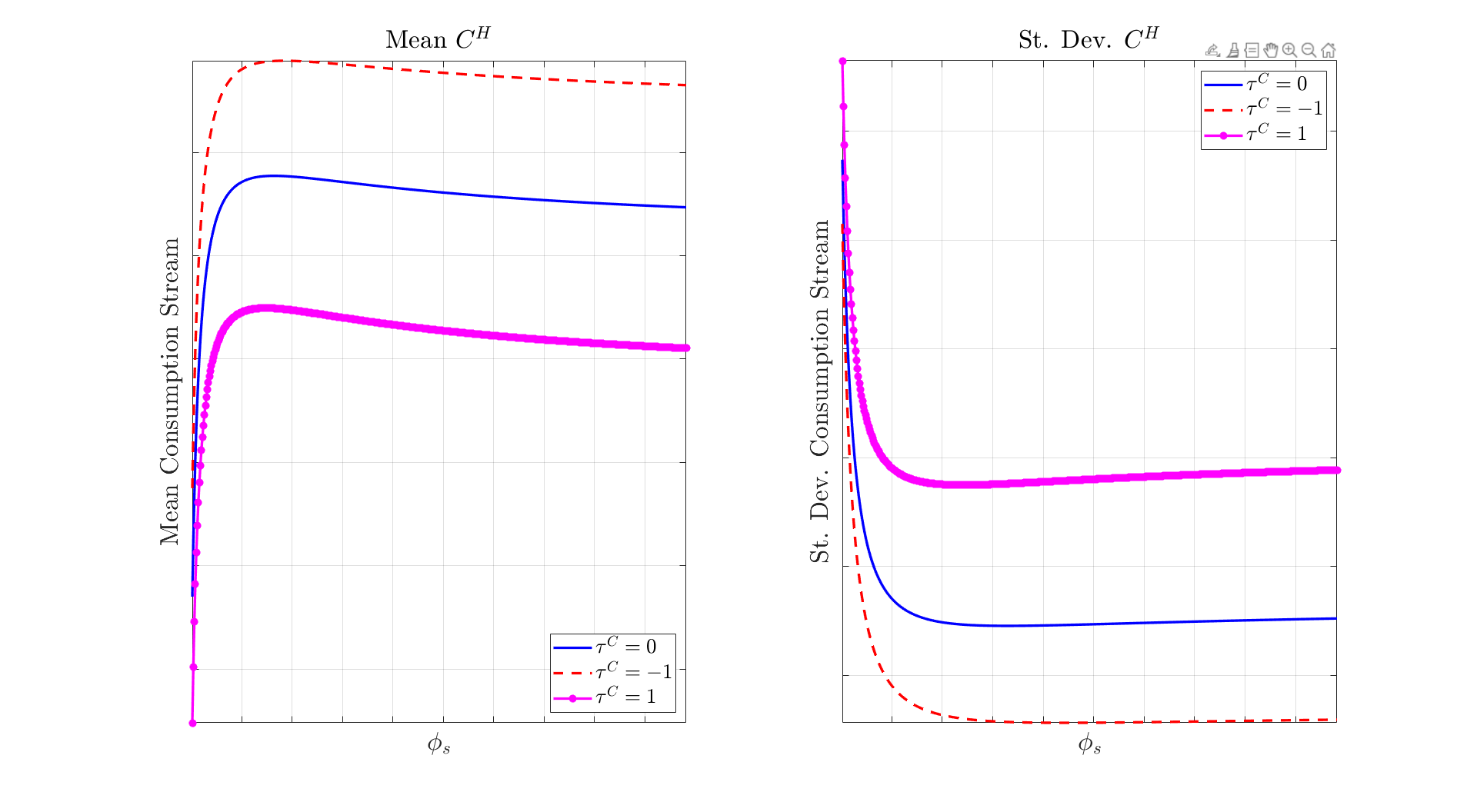}
    \floatfoot{\footnotesize \textbf{Note}: The computation exercise is carried out for the benchmark specification with non-homothetic preferences. Results are qualitatively similar for the exercise without non-homothetic preferences. }
\end{figure}

Once more, to shed light on the Hand-to-mouth households' preferences over policy regimes, Figure \ref{fig:cH_mean_stdev} shows the mean and standard deviation of consumption streams as a function of $\eta$. The first panel shows that the mean $C^H$ is decreasing with the level of fiscal pro-cyclicality.
Also, for a given fiscal policy rule, the relationship between exchange rate rigidity and mean consumption stream is U-shaped. Initial increases in exchange rate rigidity increases mean consumption streams. After certain value of $\phi_s$, further increasing exchange rate rigidity decreases mean consumption streams. However, the decreasing segment of the ``U'' is significantly less steep than the increasing segment. A similar pattern is exhibited by the standard deviation of Hand-to-mouth consumption, with volatility increasing with the degree of fiscal pro-cyclicality. For a given fiscal policy rule exchange rate rigidity has an inverted U-shape impact on consumption volatility. Again, the second segment of the ``U'' is appreciably less steep than for the Ricardian households' case. Consequently, Hand-to-mouth household's optimal policy regime implies counter-cyclical and high-degree of exchange rate rigidity. Finally, while Hand-to-mouth preferences over policy regimes are aligned with and without non-homothetic preferences, the gain in utility from the changes in mean consumption and mean labor streams of moving to the optimal policy regime is three times greater in the case with non-homothetic preferences. This is not the case for Ricardian households, in line with the results presented in Section \ref{sec:dynamics} which show that non-homothetic preferences only significantly amplify the response of Hand-to-mouth households' consumption. 

In summary, the normative analysis carried out in this section leads to the conclusion that in the current structural framework households have opposing preferences over both fiscal and monetary policy regimes. This result presents policy makers with a challenge on which policy regime to put in place. Even more, the preference of Hand-to-mouth for monetary policy regimes with significant exchange rate rigidity is a rationalization for Central Bankers' ``\textit{Fear-of-floating}'' found in the data (see \cite{calvo2002fear}). This preference over ``dirty float'' exchange rate regimes is based on the impact of exchange rate movements have on Hand-to-mouth households' consumption and labor. Given that non-homothetic preferences exacerbate the impact of exogenous shocks on Hand-to-mouth households' variables, as seen in Sections \ref{subsec:role_non_homotheticity} and \ref{subsec:redistribution_channel}, it is not surprising that the benefits of ``dirty float'' regimes are increasing in the degree of non-homothetic preferences.

\section{Conclusion} \label{sec:conclusion}

In this paper I studied the implications of household heterogeneity for the transmission of aggregate shocks and the design of optimal policy regimes. First, I use household survey data from Uruguay to highlight two dimensions of heterogeneity: access to saving technologies and exposure to commodity based goods. In particular, I show that households which report having no access to saving technologies (proxied by reporting no savings, no bank accounts or no financial assets) spend a significantly higher share of their income in commodity based nourishment than households which do have access to saving technologies. This result is in line with Engel's law, i.e., the presence of non-homothetic preferences. 

Based on this observations, I construct a Two-Agent New-Keynesian small open economy model with two key features: (i) Ricardian and hand-to-mouth agents, and (ii) non-homothetic preferences over commodity goods. I show that these features lead to aggregate responses in line with intuition introduced by \cite{alejandro1963note}. I emphasize that the presence of non-homothetic preferences exacerbate the impact of exogenous shocks, such as foreign interest rate or commodity price shocks. Finally, I argue that under this framework, domestic monetary policy has a re-distribution channel. On the one hand, an exogenous interest rate hike, appreciates the currency, reduces economic output and leads to a drop in Ricardian households' consumption. On the other hand, an exogenous interest rate hike appreciates the nominal exchange rate, reduces the domestic price of commodity based goods, leading to an increase of Hand-to-mouth household's consumption. 

Finally, I show that these features lead to households' having opposing preferences over optimal policy regimes, and provide a rationale for Central Bank's ``\textit{fear-of-floating}''. Ricardian households' prefer a more freely floating regime and a pro-cyclical fiscal rule which increase their mean consumption stream, albeit somewhat increasing its volatility. On the contrary, I show that Hand-to-mouth households prefer a ``dirty-float'' exchange rate regime and a counter-cyclical fiscal policy rule as it stabilizes the price of their consumption bundle. On the contrary, . Additionally, I argue that non-homothetic preferences increases the benefits of ``dirty-float'' exchange rate regimes (compared to an exercise without non-homothetic preferences).

\newpage
\bibliography{main.bib}

\newpage
\appendix

\section{Appendix: Additional Data Description \& Facts} \label{sec:appendix_data}

\subsection{Additional Data Description}

First, I describe the firm level data coming from Uruguay. The data employed are transaction-level customs data for the period 2001-2012. After suffering the external shock of the Argentinean 2001 sovereign default crisis, Uruguay experienced a period of rapid growth of both the economic output and total exports. Additionally, this period in which the Central Bank of Uruguay implemented an inflation targetting regime. The data was collected by the Trade and Integration Unit of the World Bank Research Department, as part of their efforts to build the Exporter Dynamics Database described in \cite{fernandes2016}.\footnote{The sources for the data for each country are detailed at \url{http://www.worldbank.org/en/research/brief/exporter-dynamics-database}.} The dataset contains information on the firm identification, value exported, quantities exported measured in kilos, HS6 good classification and the country of destination.

Additionally, across the paper I use several aggregate or macro level datasets across the paper. Data on GDP and the components of aggregate expenditure are sourced directly from the Uruguayan Central Bank.\footnote{See \url{https://www.bcu.gub.uy/Paginas/Default.aspx}. More over, the national account data is reported at the quarterly frequency for two different base years periods: 1983 and 2005. To construct a long dataset I seasonally adjust the two datasets and later splice them.} This data set is used to calibrate the structural model presented in Section \ref{sec:model} and to analyze the different policy rules presented in Section \ref{sec:optimal_policy}. I complement these aggregate datasets with international trade data which characterize . Particularly, I use the good-classification introduced by \cite{rauch1999networks} the commodity dependence indicator developed by the United Nations.\footnote{The State of Commodity Dependence report elaborated by the United Nations can be obtained at \url{https://unctad.org/webflyer/state-commodity-dependence-2019}.}

\subsection{Customs Level Findings} \label{subsec:appendix_firm_findings}

In this section of the paper I characterize the role commodity goods at the aggregate and firm level. In particular, I stress the fact that commodity goods represent the vast majority of the aggregate export bundle. Second, using firm level evidence I show that commodity goods have lower exchange rate elasticities than non-commodity or differentiated goods.

\begin{figure}[ht]
    \centering
    \caption{Share of Commodity Goods in Aggregate Exports 2019}
    \label{fig:Pie_Chart_Commodity_Horizontal}
    \includegraphics[scale=0.75]{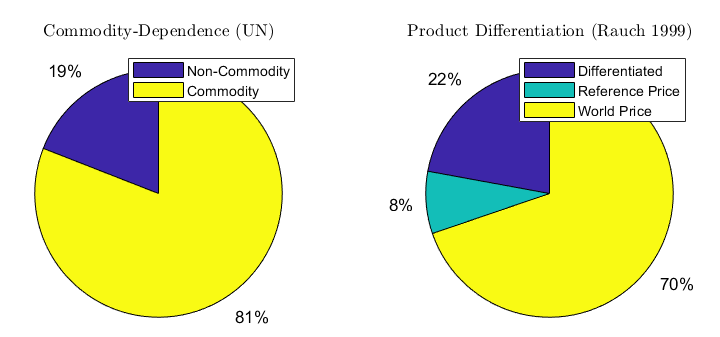}
\end{figure}

To start with, I show that agricultural commodity goods represent a significant share of the aggregate export bundle. Figure \ref{fig:Pie_Chart_Commodity_Horizontal} presents two decomposition of aggregate exports which highlight the role of commodity goods. On the left panel, I show the importance of commodity goods for Uruguay in the year 2019 according to the United Nation's commodity dependence index.\footnote{Data is sourced from ``The State of Commodity Dependence - 2019'', see \url{https://unctad.org/webflyer/state-commodity-dependence-2019}. This report is carried out every two years. The results are robust across reports.}\footnote{In more detail, goods are classified using the SITC commodity classification, Revision 3. In more detail, the codes defined as commodity goods are codes Food and live animals [0], Beverages and tobacco; [1], Crude materials, inedible, except fuels; [2], Mineral fuels, lubricants and related materials; [3], Animal and vegetable oils, fats and waxes; [4], Non-ferrous metals; [68] Pearls, precious and semi-precious stones [667], Gold, non-monetary (excluding gold ores and concentrates} Under this classification 81\% of Uruguay's total exports are commodity goods and only 19\% are non-commodity goods. 
The right panel of Figure \ref{fig:Pie_Chart_Commodity_Horizontal}, classifies Uruguay's export bundle using the good classification methodology introduced by \cite{rauch1999networks}. This classification categorizes products in goods with ``World Prices'', goods with international ``Reference Prices'', and ``Differentiated'' goods. The first and second categories are in line with concept of a commodity goods.\footnote{Products widely used and which are not meaningfully differentiated from one another.} According to this classification, 70\% of Uruguay's total exports are goods with ``World Prices'', 8\% are goods with ``Reference Prices'', and 22\% are ``Differentiated'' goods.\footnote{In \cite{rauch1999networks} introduces two good classifications: a liberal and a conservative classification. The results presented in Figure \ref{fig:Pie_Chart_Commodity_Horizontal} are constructed using the liberal good classification. When using the conservative good classification the results are: goods with ``World Prices'' 49.7\%, goods with ``Reference Prices'' 27.1\%, and ``Differentiated'' goods 23.2\%. Consequently, across the two different classifications, differentiated goods only explain a small share of Uruguay's total exports.}

Next, I turn to the estimation of exports exchange rate elasticity. To do so, I estimate the following econometric exercise
\begin{align} \label{eq:rer_elasticity}
    \ln Q_{i,p,d,t} = \mu + \beta \ln RER_{d,t} + \delta \ln GDP_{d,t} + \gamma_{i,p,d} + \Gamma_t +  \epsilon_{i,p,d,t}
\end{align}
where $Q_{i,p,d,t}$ are the exported quantities of firm $i$, of product $p$ to destination country $d$ during year $t$; $\mu$ is a constant; $RER_{d,t}$ is the bilateral real exchange rate between Uruguay and destination country $d$; $GDP_{d,t}$ is destination country $d$'s real GDP; $\gamma_{i,p,d}$ is a firm-product-destination fixed effect and $\Gamma_t$ is a year fixed effect. Under this specification, parameter $\beta$ represents the real exchange rate elasticity. This empirical specification is in line with other papers in the literature, such as \cite{li2015exchange}. In order to exclude any potential bias caused by the impact of the Argentinean 2001 debt crisis and the subsequent financial crisis, I define the start of the sample for this empirical exercise at the year 2004.
\begin{table}[ht]
    \centering
    \caption{Export Real Exchange Rate Elasticity}
    \label{tab:rer_elasticity}
    \scriptsize
\begin{tabular}{lcccccccc} \hline
 & \multicolumn{8}{c}{ $\ln Q_{i,p,d,t}$}\\
  & \multicolumn{2}{c}{Total Sample} & \multicolumn{2}{c}{World Price} & \multicolumn{2}{c}{Reference Price} & \multicolumn{2}{c}{Differentiated Goods} \\ 
    & (1) & (2) & (3) & (4) & (5) & (6) & (7) & (8) \\ \hline \hline
 &  &  &  &  &  &  &  &  \\
$\ln RER_{d,t}$ & 0.0866*** & 0.0543* & -0.00480 & -0.0524 & 0.135 & 0.119 & 0.165** & 0.137* \\
 & (0.0307) & (0.0311) & (0.0406) & (0.0416) & (0.0833) & (0.0835) & (0.0767) & (0.0782) \\
$\ln GDP_{d,t}$ &  & 0.961*** &  & 1.299*** &  & 0.572** &  & 0.991*** \\
 &  & (0.118) &  & (0.227) &  & (0.248) &  & (0.216) \\
 &  &  &  &  &  &  &  &  \\
Obs. & 51,346 & 51,346 & 11,644 & 11,644 & 9,459 & 9,459 & 21,441 & 21,441 \\
 $R^2$ & 0.934 & 0.934 & 0.898 & 0.898 & 0.908 & 0.908 & 0.921 & 0.921 \\ \hline \hline
\multicolumn{9}{c}{ Robust standard errors in parentheses} \\
\multicolumn{9}{c}{ *** p$<$0.01, ** p$<$0.05, * p$<$0.1} \\
\end{tabular}
\end{table}
Table \ref{tab:rer_elasticity} presents the results of estimating Equation \ref{eq:rer_elasticity} under different sub-samples. Columns (1) and (2) show that exports present a real exchange rate elasticity between 0.054\% and 0.087\%. for the full sample. Columns (3) - (6) show that for both ``World Price'' and ``Reference Price'' goods the real exchange rate elasticity is not significantly different from zero. However, Columns (7) and (8) show that for ``Differentiated'' goods the real exchange rate is significantly different from zero, between 0.14\% and 0.16\%. 

To test the robustness of the results presented in Table \ref{tab:rer_elasticity}, I estimate the following econometric exercise
\begin{align}
    \ln Q_{i,p,d,t} &= \mu + \beta \ln RER_{d,t} + \delta \ln GDP_{d,t} \nonumber \\ 
    & + \nu \ln RER_{d,t} \times \mathbbm{1}\left[\text{Differentiated == 1}\right] + \gamma_{i,p,d} + \Gamma_t +  \epsilon_{i,p,d,t}
\end{align}
where $\mathbbm{1}\left[\text{Differentiated == 1}\right]$ is an indicator function which takes the value of 1 if product $p$ is classified by \cite{rauch1999networks} as a differentiated good and takes the value of 0 otherwise.
\begin{table}[ht]
    \centering
    \caption{Export Real Exchange Rate Elasticity \\ Robustness-Check}
    \label{tab:rer_elasticity2}
    \footnotesize
\begin{tabular}{lccc} \hline
 & \multicolumn{3}{c}{$\ln Q_{i,p,d,t}$} \\ 
  & (1) & (2) & (3) \\ \hline \hline
 &  &  &  \\
$\ln RER_{d,t}$ & 0.0866*** & 0.0282 & -0.00829 \\
 & (0.0307) & (0.0325) & (0.0329) \\
$\ln RER_{d,t} \times \mathbbm{1}\left[\text{Differentiated == 1}\right]$ &  & 0.307*** & 0.325*** \\
 &  & (0.0640) & (0.0648) \\
$\ln GDP_{d,t}$ &  &  & 0.981*** \\
 &  &  & (0.118) \\
Constant & 8.706*** & 8.588*** & 1.453* \\
 & (0.0236) & (0.0365) & (0.857) \\
 &  &  &  \\
Obs & 51,346 & 51,346 & 51,346 \\
$R^2$ & 0.934 & 0.934 & 0.934 \\ \hline \hline
\multicolumn{4}{c}{ Robust standard errors in parentheses} \\
\multicolumn{4}{c}{ *** p$<$0.01, ** p$<$0.05, * p$<$0.1} \\
\end{tabular}
\end{table}
Consequently, parameter $\beta$ represents the real exchange rate elasticity of exports of ``World Price'' and ``Reference Price'' goods; while the sum of $\beta$ and $\nu$ represents the real exchange rate elasticity of exports of ``Differentiated'' goods. Table \ref{tab:rer_elasticity2} presents the results of estimating this econometric exercise. In line with the results presented in Table \ref{tab:rer_elasticity}, the real exchange rate elasticity seems to be non-significantly different from zero for the ``World Price'' or ``Reference Price'' goods, while it is significantly different from zero for ``Differentiated'' goods. 

Finally, I turn to estimating the relative prices of commodity and non-commodity goods using customs level data. This relative price is of importance for the calibration of key steady state variables of the model (carried out in Section \ref{subsec:calibration}. To do so, I compute the median price of export goods under the three categories of the \cite{rauch1999networks} good classification for each year in the sample. 
\begin{figure}[ht]
    \centering
    \caption{Evolution of the Relative Price of Commodity Goods}
    \label{fig:Median_Prices}
    \includegraphics[scale=0.6]{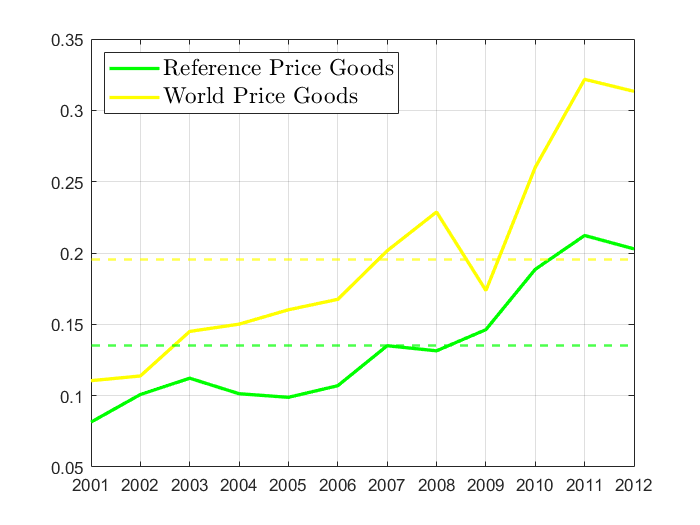}
    \floatfoot{\textbf{Note}: Relative prices are computed using customs level data. These prices are computed as the ratio of the yearly median price of ``Reference Price'' and ``World Price'' goods divided by the median price of ``Differentiated'' goods. }
\end{figure}
The relative price of both ``Reference Price'' and ``World Price'' goods exhibit a clear upper trend during the time sample, in line with the commodity super cycle experienced between 2003 and 2014. The relative prices of these goods with ``Differentiated'' goods fluctuates between 0.08 in the year 2001 and 0.32 in the year 2012. Furthermore, the sample averages of the relative prices are 0.13 and 0.19 for ``Reference Price'' and ``World Price'', respectively. Consequently, I set the steady state value of $\bar{P}^{Co}$ to 0.15. 

\subsection{Evidence on the International \& Domestic Price of Food} \label{subsec:appendix_price_food}

In this section of the Appendix I present evidence on the relationship between domestic and international prices of food goods. The model presented in Section \ref{sec:model} assumes that the domestic price of the commodity good is equal to the international price divided by the nominal exchange rate. In order words, I assume that the law of one price is satisfied in the model. While the literature in international economics has challenged the law of one price, some evidence suggests that it still applies to commodity goods (see \cite{baffes1991some}). I argue that the 

First, I describe the datasets used in this analysis. I use the consumer price index information from the INE or National Institute of Statistics of Uruguay for the period January 2001 to December 2010.\footnote{Data is available at \url{https://www.ine.gub.uy/ipc-indice-de-precios-del-consumo}} Consolidated time indexes are not available, and consequently this is the index in roughly the same time frame as used on the other empirical analysis carried out in the paper. In particular, I use the sub-chapter of the index that focuses on food and nourishment goods. Additionally, I use the nominal exchange rate of the Uruguayan peso with respect to the US dollar sourced from the IMF. Lastly, I use the International Monetary Fund's ``Global price of Food Index'', constructed as the a value representing the benchmark prices which are representative of the global market, determined by the largest exporter of a given commodity. (prices computed as the period averages in nominal U.S. dollars).\footnote{Note that this international food price index is based on commodity goods' prices in global markets. Thus, it does not fully represent the international price at which Uruguay could purchase these commodity goods.}

In order to test how close the domestic and international prices of food move I compute the correlation between the international price of food, the domestic price of food, the nominal exchange rate, and the dollar price of food (i.e., domestic price of food divided by the nominal exchange rate). 
\begin{table}[ht]
    \centering
    \caption{Domestic and International Food Prices \\ \small Correlation Analysis}
    \label{tab:correlation_food}
    \small
    \begin{tabular}{l c c c c}
    & Int. Price & Price in Pesos & Price in Dollars & NER  \\ \hline \hline
    \multicolumn{4}{c}{ } \\
    Corr. in Levels & 1 & 0.9120*** & 0.8307*** &  -0.0150 \\
    \multicolumn{4}{c}{ } \\
    Corr. in Rates & 1 & 0.3856** & 0.3894** & -0.1931 \\
    \multicolumn{4}{c}{ } \\\hline \hline
    \end{tabular}
    \floatfoot{\small \textbf{Note:} The correlations are computed at the quarterly level with respect to the international food price index. This is in line with the parametrization of the model described in Section \ref{subsec:calibration}. The growth rates are computed as the log difference of the variables with respect to the value from the same quarter of the previous year. The stars denote different levels of statistical significance:  *** p$<$0.01, ** p$<$0.05, * p$<$0.1. }
\end{table}
The first row of Table \ref{tab:correlation_food} presents the correlation between the international food price index and domestic variables in levels. On the one hand, both the domestic peso and dollar price index of food are highly correlated with the international price, 0.91 and 0.83 respectively. On the other hand, the nominal exchange rate exhibits no significant correlation. However, these variables may exhibit a unit root which would imply that the correlation captured is spurious. Consequently, I compute the correlation between the inter-annual growth rates of these variables, presented in the second row of Table \ref{tab:correlation_food}. While the computed correlations between the international and domestic prices of food are lower, they are still significantly different from zero. 

In addition to the preceding correlation analysis, I carry out a co-integration analysis. More concretely, I follow \cite{johansen1995likelihood} and estimate the test of co-integration between the log of the international food price, the domestic peso food price and the nominal exchange rate. One would expect that there is one co-integration relationship between these variables as the international price of food expressed in US dollars should, in the long run, track the domestic price of food in US dollars. 
\begin{table}[ht]
    \centering
    \caption{Domestic and International Food Prices \\ \small Co-integration Analysis}
    \label{tab:cointegration_food}
    \begin{tabular}{c c c}
        Maximum Rank &  Trace Statistics & 5\% Critical Value \\ \hline \hline
        \multicolumn{3}{c}{ } \\
        0 & 36.7742 & 29.68 \\
        1 & \textbf{12.6547*} & 15.41 \\
        2 & 0.0941 & 3.76 \\
        \multicolumn{3}{c}{ } \\ \hline \hline
    \end{tabular}
    \floatfoot{\small \textbf{Note:} The Johansen co-integration test is carried out using Stata. The time sample is 2001-Q1 to 2010-Q4. The test is estimated using a model with 2 lags. The results are robust to using 4 lags.}
\end{table}
The results are presented in Table \ref{tab:cointegration_food}, which reject the null hypothesis of no co-integrating equations. In addition, the data suggests that there is no more than one co-integrating equations. Thus, I take this empirical result as further evidence of a strong relationship between international and domestic food prices. 

In conclusion, in the present section of the appendix I showed that there is a strong correlation between international and domestic food price indexes. Furthermore, I show that the data suggests that there is too a long run relationship between these variables. I take this empirical results as evidence that the modelling assumptions used in the structural framework in Section \ref{sec:model} are justified given the goals of the present paper. 




\section{Appendix: Model Details} \label{sec:appendix_model}

\subsection{Firms} \label{subsec:model_firms}

Next, I turn to describing the production side of the economy. This side and its features are standard to small open economy New Keynesian models (see \cite{christiano2011introducing}). First, there is a block of intermediate and final good producing firms subject to pricing frictions and working capital constraints which produce the domestic homogeneous good and introduce New Keynesian features to the model. Second, there is a block of investment and capital good producing firms subject to investment adjustment costs. Third, there is a block of exporting firms comprised of commodity and non-commodity exporters.

\subsubsection{Domestic Homogeneous Good} \label{subsubsec:domestic_homogeneous_good}

The production of the homogeneous good can be divided into two stages: the intermediate good stage and the final good stage. The domestic final good is produced via a CES aggregation of intermediate domestic goods:
\begin{equation*}
    Y_t = \left[ \int^1_0 Y^{\frac{\epsilon-1}{\epsilon}}_{i,t} di \right]^{\frac{\epsilon}{\epsilon-1}} 
\end{equation*}
where $\epsilon$ is a parameter which governs the elasticity of substitution between the intermediate good varieties and is assumed to be greater than one, $\epsilon>1$, and $Y_{i,t}$ represents the amount used of the domestic intermediate good $i$. This framework leads to the following demand schedules for each variety $i$
\begin{equation*}
    Y_{i,t} = Y_t \left(\frac{P_t}{P_{i,t}}\right)^{\epsilon}
\end{equation*}
where $P_t = \left( \int^1_0 P^{1-\epsilon}_{i,t} di \right)^{\frac{1}{1-\epsilon}}$.

There is a continuum of intermediate good firms of measure one which produce combining capital and labor using a standard Cobb Douglas production function
\begin{equation*}
    Y_{i,t} = K^{\alpha}_{i,t-1} \left(A_t l_{i,t}\right)^{1-\alpha}
\end{equation*}
where $A_t$ is a technology process whose growth rate follows an AR(1) process.\footnote{The presence of this technology process which grows in time implies that in order to solve this model, I need to deflate real variables.} The terms $l_{i,t}$ and $K_{i,t-1}$ are the amounts of total labor and capital services hired in period $t$.\footnote{Notation is slightly confusing for capital services. Note that $K_{i,t-1}$ represents the demand by firm $i$ at time $t$. I choose a notation of $t-1$ to make sure that when the equilibrium conditions are established, market clearing conditions are correctly specified as the supply of capital is determined by households in period $t-1$.} 

Firms are subject to standard Calvo price-setting frictions. This implies that with probability $1-\theta$ the firm is able to reset its price and with probability $\theta$ it has to keep its price from period $t-1$. Firm $i$'s marginal cost can be written as
\begin{equation*}
    MC_t = \frac{\left(1-\nu\right) W_t / P_t }{\left(1-\alpha \right) \left(\frac{K_{t-1}}{l_{t}}\right)^{\alpha} A^{1-\alpha}_t}
\end{equation*}
where $\nu$ represents a (potential) subsidy to employment and  $W_t$ represents the nominal aggregate cost of the labor bundle. Note that in the expression for the marginal cost the use of capital and labor are expressed without the $i^{th}$ subscript. This is because as firms behave atomistically, they all face the same prices independent of $i$. The firm's price-setting objective function can be written as
\begin{equation*}
    \mathbb{E}^{i}_t \sum^{\infty}_{j=0} \beta^j v_{t+j} \left[P_{i,t+j} Y_{i,t+j} - P_{t+j} s_{t+j} Y_{i,t+j} \right]
\end{equation*}
where $v_{t+j} = u'\left(C_{t+j}\right) / P^c_{t+j}$ is the marginal value of an additional unit of currency or revenue for firm $i$. Also, the subscript in the expectation implies that it involves the idiosyncratic Calvo shock. 

The solution to firm $i$'s optimal price-setting problem leads to three  equilibrium conditions that relate inflation to marginal costs
\begin{align}
    \mathcal{K}_t &= \frac{1}{p^c_t} \frac{y_t}{c^{R}_t} \frac{\epsilon}{\epsilon-1}MC_t + \beta \theta \mathbb{E}_t \pi^{\epsilon}_{t+1} \mathcal{K}_{t+1} \\
    \mathcal{F}_t &= \frac{y_t}{p^c_t c^R_t} + \beta \theta \mathbb{E}_t \pi^{\epsilon-1}_t \mathcal{F}_{t+1} \\
    \frac{\mathcal{K}_t}{\mathcal{F}_t } &= \left[\frac{1-\theta \pi^{\epsilon-1}_t}{1-\theta} \right]^{\frac{1}{1-\epsilon}}
\end{align}
Aggregate output of the domestic final good is given by
\begin{equation*}
    Y_t = p^{*}_t K^{\alpha}_{t-1} \left(A_t l_t \right)^{1-\alpha}
\end{equation*}
where $p^{*}_t$ is a measure of the distortion arising from price dispersion by price-setting frictions. The law of motion for this distortion is given by 
\begin{equation}
    p^{*}_t = \left[ \left(1-\theta\right) \left( \frac{1-\theta \pi^{\epsilon-1}_t}{1-\theta} \right)^{\frac{\epsilon}{\epsilon-1}} + \frac{\theta \pi^{\epsilon}_t}{p^{*}_{t-1}} \right]
\end{equation}


\subsubsection{Capital Goods and Investment} \label{subsec:capital_goods}

There is a large and constant number of identical capital-producing firms which act competitively (i.e., take prices of inputs and outputs as given). Each of these firms produce capital by combining old capital with investment goods.

To produce investment goods, $I_t$, the capital good firms combine domestic and foreign investment goods using a CES technology
\begin{equation*}
    I_t = \left[\gamma^{\frac{1}{\nu_I}}_I \left(I_{d,t}\right)^{\frac{\nu_I-1}{\nu_I}} + (1-\gamma_I)^{\frac{1}{\nu_I}} \left(I_{f,t}\right)^{\frac{\nu_I-1}{\nu_I}} \right]^{\frac{\nu_I}{\nu_I-1}}
\end{equation*}
I assume that the domestic and foreign investment goods are the domestic and foreign homogeneous goods, which lead to $P_{I,d,t} =P_t$, $P_{I,f,t} =S_t P^f_t$. Next, every individual capital producer accumulates capital from period $t-1$ to period $t$ using the following law of motion:
\begin{equation*}
    K_{t} = K_{t-1} \left(1-\delta^K \right) + \left[1 - S\left(\frac{I_t}{I_{t-1}}\right)  \right]I_t
\end{equation*}
where $\delta^K$ is the depreciation rate of capital and function $S(.)$ is a convex function that introduces adjustment costs of investment. I assume $S(g_I) = \frac{\kappa^I}{2}\left(g_I - \mu_I\right)^2$, where $g_I = \frac{I_t}{I_{-1}}$.

\subsubsection{Exports} \label{subsubsec:exports}

This section describes the exporting side of the economy. In order to match the dynamics of a small open economy similar to those of Latin America, exports are composed of two distinct sub-sectors: a commodity sector and a non-commodity sector. On one hand, the non-commodity exporting sector faces a downward sloping demand curve which depend on a foreign demand shifter and the real exchange rate (in line with standard SOE-NK models). On the other hand, the commodity sector is characterized by stochastic processes which govern the international price of the good and the local endowment the commodity good.

\noindent
\textbf{Non-Commodity Exports.} There is a continuum of non-commodity exporting firms. The amount exported of non-commodity goods is denoted by $X_t$. Foreign demand for this good is given by 
\begin{equation} 
    X_t = \left(\frac{P^x_t}{P^f_t} \right)^{-\eta_f} Y^f_t
\end{equation}
where $P^x_t$ is the price in foreign currency of the exported good, $Y^f_t$ is a foreign demand shifter and $\eta_f$ is a parameter which governs demand elasticity. Foreign demand is exogenous to the domestic economy,
\begin{equation*}
    Y^f_t = y^f_t Z_t
\end{equation*}
where $Z_t$ grows with $A_t$, yet $Z_t$ responds extremely sluggishly to changes in $A_t$. I follow \cite{christiano2011introducing}, and choose the following law of motion for $Z_t$
\begin{equation*}
    Z_t = A^{1-\delta}_t Z^{\delta}_{t-1}, \quad 0 < \delta < 1
\end{equation*}
where 
\begin{equation*}
    z_t \equiv \frac{Z_t}{A_t} = \exp \left( - \delta \Delta a_t \right) z^{\delta}_{t-1}
\end{equation*}
Note that $Z_{t}$ grows at the same rate as $A_{t}$ in the sense that $Z_{t}/A_{t}$ converges to a constant in steady state. With $\delta$ close, but less than 1, $Z_{t}$ hardly responds to a shock in $A_{t}$. In the above expression, $p_{t}^{x}$ is the terms of trade: 
\begin{equation} 
    p_{t}^{x}=\frac{P_{t}^{x}}{P_{t}^{f}}=\frac{P_{t}}{S_{t}P_{t}^{f}}=\frac{P_{t}}{P_{t}^{c,R}}\frac{P_{t}^{c,R}}{S_{t}P_{t}^{f}}=\frac{1}{p_{t}^{c,R}q_{t}}=\frac{1}{p_{t}^{m}}    
\end{equation}
where $p^{c,R}_t$ is the relative price of the consumption bundle of the Ricardian household with respect to the homogeneous final good, and $q_{t}$ denotes the real exchange rate,
\begin{align} 
    q_{t}=\frac{S_{t}P_{t}^{f}}{P_{t}^{c,R}}    
\end{align}

\noindent
\textbf{Commodity Exports.} The commodity side of exports is governed by stochastic processes. Both prices and quantities produced of the commodity good are determined by exogenous processes. In particular, I assume that these processes follow an $AR(1)$ process in logarithms
\begin{align}
    \log \left(\frac{y^{C}_t}{\bar{y^C}}\right) &= \rho^C \log \left(\frac{y^{C}_{t-1}}{\bar{y^C}}\right) + \epsilon^C_t \\ 
    \log \left(\frac{P^{Co}_t}{\Bar{P^C}}\right) &= \rho^P \log \left(\frac{P^{C}_{t-1}}{\Bar{P^{Co}}}\right) + \epsilon^P_t 
\end{align}
where $y^{C}_t$ and $\bar{y}^{C}$ are deflated amounts produced of the commodity good in period $t$ and in steady state; $P^{C}_t$ and $\bar{P^{C}}$ are the prices of the commodity good in period $t$ and in steady state expressed in foreign currency; and $\epsilon^{C}_t$ and $\epsilon^{P}_t$ are Gaussian innovations with standard deviations given by $\sigma^{C}_t$ and $\sigma^{P}_t$.\footnote{The assumption of an stochastic endowment of the commodity good is a simplification but describes in a reduced form way that commodity production, like soybeans, wheat and other agricultural products are subject to shocks external to the business cycle, such as droughts or floods, and intensive in natural capital. Furthermore, given the nature of agricultural production, once the sowing process is done, higher levels of employment will not increment production significantly. }

The exported quantities of the sector are determined by the difference between the domestic production  and the local consumption of the commodity good
\begin{align*}
    X^{Co}_t &= Y^{Co}_t - C^{Co}_t \\
    X^{Co}_t &= Y^{Co}_t - \lambda^R C^{R,Co}_t - \left(1-\lambda^R\right) C^{H,Co}_t
\end{align*}
In the parametrization of the model in Section \ref{subsec:calibration}, parameters are chosen such that the economy is a net-exporter of the commodity good both in steady state and for reasonable values of the stochastic processes in the economy.\footnote{In other words, parameters are chosen such that in both the non-stochastic steady state of the economy and under simulations the trade balance of the commodity good is positive.}

\subsection{Aggregation \& Market Clearing } \label{subsec:market_clearing}

There are three main aggregate market clearing conditions. First, we have the economy-wide budget constraint which relates the trade balance and financial flows into and out the economy
\begin{align}
    \underbrace{\frac{P_t}{S_t} X_t + P^{Co}_t\left(Y^{Co}_t - C^{R,Co}_t - C^{H,Co}_t \right)}_{\text{Exports}}  - \underbrace{P^f_t\left[ I_{m,t} + C^R_{m,t} + C^H_{m,t}\right]}_{\text{Imports}} \nonumber \\
    \qquad =-\underbrace{\left(F^O_t-R^{*}_{t-1}F^O_{t-1}\right)}_{\text{Net Foreign Asset Accumulation}}
\end{align}
where $I_{m,t}, C^{R}_{m,t}$ and $C^{H}_{m,t}$ represent the imports to carry out investment, consumption by the Ricardian household and the Hand-to-Mouth household, respectively. Note that, as Ricardian households are the only agents in the economy which have access to international financial markets, the net foreign asset accumulation must be equal to this households' change in foreign currency assets.

Next, there is a market clearing condition for the domestic homogeneous final good
\begin{align*}
    Y_t = I_{d,t}  + \lambda^R C^{R,D}_t + \left(1-\lambda^{R}\right) C^{H,D}_t +G_t + X_t 
\end{align*}
where $I_{d,t}$ represents the amount of investment carried out using the domestic homogeneous good, $C^{R,D}_t$ and $C^{H,D}_t$ represents the consumption of the domestic good by each type of household, $G_t$ is public expenditure, and $X_t$ represents non-commodity exports.

Finally, labor markets have to clear, such that the total amount of labor employed in the production of the domestic homogeneous good has to be equal to the sum of the labor supplied by each type of household
\begin{align*}
    l_t = \lambda^R_t l^R_t + \left(1-\lambda^R\right) l^H_t
\end{align*}

\subsection{Equilibrium Conditions} \label{subsec:appendix_equilibrium_conditions}

In this appendix I present the model's full equilibrium conditions. Given that the model has long run productivity growth I deflate real variables by dividing by the previous period technology level, i.e. $A_{t-1}$. Furthermore, given that the economy exhibits a positive steady state inflation rate, prices are expressed as relative with respect to the domestic final good.

\noindent
\textbf{Calvo Pricing:} Five (5) pricing equations
\begin{align}
    \mathcal{K}_t &= \frac{1}{p^c_t} \frac{y_t}{c^{R}_t} \frac{\epsilon}{\epsilon-1}MC_t + \beta \theta \mathbb{E}_t \pi^{\epsilon}_{t+1} \mathcal{K}_{t+1} \\
    \mathcal{F}_t &= \frac{y_t}{p^c_t c^R_t} + \beta \theta \mathbb{E}_t \pi^{\epsilon-1}_t \mathcal{F}_{t+1} \\
    \frac{\mathcal{K}_t}{\mathcal{F}_t } &= \left[\frac{1-\theta \pi^{\epsilon-1}_t}{1-\theta} \right]^{\frac{1}{1-\epsilon}} \\
    p^{*}_t &= \left[ \left(1-\theta\right) \left( \frac{1-\theta \pi^{\epsilon-1}_t}{1-\theta} \right)^{\frac{\epsilon}{\epsilon-1}} + \frac{\theta \pi^{\epsilon}_t}{p^{*}_{t-1}} \right] \\
    MC_t &= (1-\nu)  \exp\left(\alpha \times \Delta A_t \right) \frac{w_t}{\left((1-\alpha) \times k_{t-1}/l_t^{\alpha} \right)}
\end{align}

\noindent
\textbf{Ricardian Household's Euler Equations:} Three (3) equations
\begin{align}
    1 &= \beta \mathbb{E} \left[ \frac{c^R_{t}}{ \exp \left( \Delta A_{t+1}\right) c^R_{t+1} } \frac{R_{d,t}}{\pi^{C,R}_{t+1}} \right] \\ 
    \frac{1}{c^R_t} &= -\gamma \left(\frac{d^{*}_t}{z_t p^{C,R}_t} - \Upsilon_t\right) + \beta \mathbb{E} \left[ \frac{s_{t+1} R^{*}_{t}}{ \pi^{C,R}_{t+1} c^{R}_{t+1} \exp \left( \Delta A_{t+1}\right) } \right] \\
    1 &= \beta \mathbb{E} \left[ \frac{c^R_{t}}{ \exp \left( \Delta A_{t+1}\right) c^R_{t+1} } \frac{R^{k}_{t+1}}{\pi^{C,R}_{t+1}} \right]
\end{align}

\noindent
\textbf{Final Good Production \& Market Clearing conditions:} Three (3) equations. Final goods production and market clearing, non-commodity aggregator good, 
\begin{align}  
    y_{t}&=p_{t}^{*}\left(\frac{k_{t-1}}{\exp\left(\varDelta a_{t}\right)}\right)^{\alpha}l_{t}^{1-\alpha} \\
    y_t &= \gamma_I \left(p^{I}_t\right)^{\nu_I} i_t+\lambda^R (1-\omega^D)(p^{c,R})^{\eta_c}c^{N,R}+(1-\lambda^R)(1-\omega^D)\left(p^{c,H}\right)^{\eta_c}c^{N,H}_t+x_t+g_t z_t \\
    c^{N}_t &= \lambda^R c^{N,R}_t + (1-\lambda^R)*c^{N,H}_t
\end{align}

\noindent
\textbf{Relationship across prices:} Five (5) equations
\begin{align}
      p^N_t &= \left[\left(1-\omega^D\right) + \omega^D \left(p^m_t\right)^{1-\eta_c}\right]^{1/(1-\eta_c)} \\  
      p^m_t &= p^{c,R}q_t \\ 
      p^{Co}_t  &= s_t P^{Co}_t \\
      p^{c,R}_t &= \left(p^{Co}/\alpha_{Co}\right)^{\alpha_{Co}}  \left(p^N_t/(1-\alpha_{Co})\right)^{1-\alpha_Co} + \frac{p^{Co}_t\phi_{Co}}{c^{R}_t} \\
      p^{c,H}_t &= \left(p^{Co}/\alpha_{Co}\right)^{\alpha_{Co}}  \left(p^N_t/(1-\alpha_{Co})\right)^{1-\alpha_Co} + \frac{p^{Co}_t\phi_{Co}}{c^{H}_t}
\end{align}

\noindent
\textbf{Relationship across inflation rates:} Four (4) equations
\begin{align}
    \pi^N_t  &= \pi_t  p^N_t/p^N_{t-1} \\
    \pi^{Co} &= \pi_t  p^{Co}_t/p^{Co}_{t-1} \\
    \pi^{c,R}_t &= \pi_t p^{c,R}_t/p^{c,R}_{t-1} \\
    \pi^{c,H}_t &= \pi_t p^{c,H}_t/p^{c,H}_{t-1}
\end{align}

\noindent
\textbf{Ricardian Household's consumption conditions} Four (4) equations
\begin{align}
    C^{Co,R}_t &= \frac{\alpha_{Co} C^R_t}{S_t P^{Co}_t \left( \frac{\alpha_{Co}}{S_t P^{Co}_t} \right)^{\alpha_{Co}} \left( \frac{1-\alpha_{Co}}{P^{N}_t} \right)^{1-\alpha_{Co}} } + \phi_{Co} \\
    C^{N,R}_t &= \frac{\left(1-\alpha_{Co}\right) C^R_t}{P^{N}_t \left( \frac{\alpha_{Co}}{S_t P^{Co}_t} \right)^{\alpha_{Co}} \left( \frac{1-\alpha_{Co}}{P^{N}_t} \right)^{1-\alpha_{Co}}} \\
    C^{D,R}_t &= \omega^{D}_c \left(\frac{P^{N}_t}{P_t} \right)^{\eta_c} C^{N,R}_t \\
    C^{F,R}_t &= \left(1-\omega^{D}_c\right) \left(\frac{P^{N}_t}{S_t P^F_t} \right)^{\eta_c} C^{N,R}_t 
\end{align}

\noindent
\textbf{Hand-to-mouth Household's consumption conditions}  Four (4) equations
\begin{align}
    C^{Co,H}_t &= \frac{\alpha_{Co} C^H_t}{S_t P^{Co}_t \left( \frac{\alpha_{Co}}{S_t P^{Co}_t} \right)^{\alpha_{Co}} \left( \frac{1-\alpha_{Co}}{P^{N}_t} \right)^{1-\alpha_{Co}} } + \phi_{Co} \\
    C^{N,H}_t &= \frac{\left(1-\alpha_{Co}\right) C^H_t}{P^{N}_t \left( \frac{\alpha_{Co}}{S_t P^{Co}_t} \right)^{\alpha_{Co}} \left( \frac{1-\alpha_{Co}}{P^{N}_t} \right)^{1-\alpha_{Co}}} \\
    C^{D,H}_t &= \omega^{D}_c \left(\frac{P^{N}_t}{P_t} \right)^{\eta_c} C^{N,H}_t \\
    C^{F,H}_t &= \left(1-\omega^{D}_c\right) \left(\frac{P^{N}_t}{S_t P^F_t} \right)^{\eta_c} C^{N,H}_t 
\end{align}

\noindent
\textbf{Nominal and Real Exchange Rate Conditions:} Three (3) equations
\begin{align}
1 &= q_t p^x_t p^{c,R}_t \\
q_t/q_{t-1} &= s_t  \pi^f/\pi^{c,R} \\
s_t &= \psi \exp\left(\tilde{S}_t-\tilde{S}_{t-1}\right) 
\end{align}

\noindent
\textbf{Investment conditions.} Five (5) equations
\begin{align}
    p_{t}^{k} &=\frac{p_{I,t}}{1-S\left(\frac{i_{t}}{i_{t-1}}\exp\left(\varDelta A_{t}\right)\right)-S'\left(\frac{i_{t}}{i_{t-1}}\exp\left(\varDelta A_{t}\right)\right)\frac{i_{t}}{i_{t-1}}\exp\left(\varDelta A_{t}\right)} \nonumber \\
	& -\beta\frac{c_{t}}{\pi_{t+1}^{c}c_{t+1}\exp\left(\varDelta A_{t+1}\right)}\times\frac{\pi_{t+1}p_{t+1}^{k}S'\left(\frac{i_{t+1}}{i_{t}}\exp\left(\varDelta A_{t+1}\right)\right)\left(\frac{i_{t+1}}{i_{t}}\exp\left(\varDelta A_{t+1}\right)\right)^{2}}{1-S\left(\frac{i_{t}}{i_{t-1}}\exp\left(\varDelta A_{t}\right)\right)-S'\left(\frac{i_{t}}{i_{t-1}}\exp\left(\varDelta A_{t}\right)\right)\frac{i_{t}}{i_{t-1}}\exp\left(\varDelta A_{t}\right)} \\ 
	k_{t}& =\left(1-\delta\right)\exp\left(-\varDelta A_{t}\right)k_{t-1}+\left[1-S\left(\frac{i_{t}}{i_{t-1}}\exp\left(\varDelta A_{t}\right)\right)\right]i_{t}  \\
	r_t &= \frac{\alpha}{1-\alpha}  \frac{ w_t \alpha  \exp(\varDelta A_t)}{(k_{t-1}} \\
	R^k_t &= \pi_t \frac{\left(r_t+(1-\delta_K\right)p^k_t}{p^k_{t-1}} \\
    p^{I}_t &= \left[\gamma_I+(1-\gamma_I)\left(p^m_t\right)^{1-\nu_I}\right]^{1/(1-\nu_I)}
\end{align}

\noindent
\textbf{Households' and Market Labor Conditions:} Four (4) equations
\begin{align}
    l_t &= \lambda^R l^R_t + (1-\lambda^R) l^H_t \\
    w_t &= p^{c,H}_t c^H_t (l^H_t)^{\varphi} \chi_H \\
    w_t &= p^{c,R}_t c^R_t (l^R_t)^{\varphi} \chi_R \\
    w_t &= (1-\alpha)\left(k_{t-1}\right))^{\alpha} l^{-\alpha}_t \exp(\varDelta A_t)^{1-\alpha}  
\end{align}

\noindent
\textbf{Scale factor:} One (1) equation
\begin{align}
    z_t = \exp\left(-\delta \varDelta A_t\right)  z_{t-1}^{\delta}
\end{align}

\noindent
\textbf{External Balance Conditions:} Two (2) equations
\begin{align}
    & P^{Co}\left(Y^{Co}_t - \lambda^R c^{Co,R}_t - \left(1-\lambda^R\right) c^{Co,H}_t \right)  \left(p_{t}^{x}\right)^{\eta_{f}}y_{t}^{f}z_{t} + \nonumber \\
    & -p_{t}^{m}\left(\left(1-\gamma_{I}\right)\left(\frac{p_{I,t}}{p_{t}^{m}}\right)^{\nu_{I}}i_{t}+\lambda^R\omega^{D}\left(\frac{p_{t}^{N}}{p_{t}^{m}}\right)^{\eta_{c}}c^{N,R}_{t} +    \left(1-\lambda^R\right) \omega^{D}\left(\frac{p_{t}^{N}}{p_{t}^{m}}\right)^{\eta_{c}}c^{N,H}_{t} \right) \nonumber \\
    &=-\left(f_{t}^{o}-\frac{s_{t}R_{t-1}^{*}}{\pi_{t}e^{\varDelta A_{t}}}f_{t-1}^{o}\right) \\
    d^{*}_t &= f^{O}_t
\end{align}

\noindent
\textbf{Policy Equations.} Two (2) equations
\begin{align}
    \log \left(\frac{R_{d,t}}{\bar{R}_d}\right) &= \rho_R \log  \left(\frac{R_{d,t-1}}{\bar{R}_d}\right) \nonumber \\
    & \quad + \left(1-\rho_R\right) \left[ \phi_{\pi} \log\left(\frac{\lambda^R\pi^{C,R}_{t} + \left(1-\right)\pi^{C,H}_t}{\bar{\pi}} \right) + \phi_y \log\left(\frac{y_{t}}{\bar{y}} \right) + \phi_s \log \left(\tilde{S}_t\right) \right] + \epsilon^R_t \\
    G_t &= \bar{g} + \tau^C \left(\frac{P^{Co}_t Y^{Co}_t}{\bar{P^{Co}}\bar{Y}^{Co}} -1 \right) 
\end{align}

\subsection{Description Steady State Computation} \label{subsec:appendix_model_steady_state}

In this appendix I briefly describe the computation of this economy's steady state. The equilibrium conditions described above are boiled down to two equations two unknowns. The equations are organized into an `inner loop, outer loop' structure. In the outer loop we search over the values of one model endogenous variable, I call it $\tilde{\varphi}$, to satisfy one steady state condition. The inner loop variable is $R^{k}$. Conditional on a value of $\tilde{\varphi}$, $R^{k}$ solves one equation in just $R^{k}$. 

The best way to go is to create a grid of values of $x=\tilde{\varphi}$ and $y=R^{k}$. The two equations to be solved are:
\begin{align*}
f\left(x,y\right)	&=0 \\
g\left(x,y\right)	&=0    
\end{align*}
For each $x$ solve for $y\left(x\right)$ which satisfies
\begin{align*}
    f\left(x,y\left(x\right)\right)=0   
\end{align*}
Search over $x$ so that
\begin{align*}
    g\left(x,y\left(x\right)\right)=0   
\end{align*}
The only thing being solved here is a one dimensional problem, 
\begin{align*}
    h\left(z\right)=0   
\end{align*}
To solve, put a fine grid on $z$, $\left[z_{1},...,z_{N}\right]$. Then, evaluate $h\left(z_{i}\right)$ for $i=1,...,N$. Verify there is only one sign switch in $h\left(z_{i}\right)$. Suppose $h\left(z_{\tilde{i}}\right)h\left(z_{\tilde{i}-1}\right)<0$. Provide $\left[z_{\tilde{i}-1},z_{\tilde{i}}\right]$ to some type of function such as ``fzero.m'' in MatLab and let it choose $z$ to drive $h$ to zero. This procedure ensure that if there is a unique steady state, I found it.

\newpage
\section{Appendix: Additional Parametrization Details} \label{sec:appendix_calibration}

In this appendix I present the numerical values for the rest of the model's parameters, which have not been described in Tables \ref{tab:micro_calibration} and \ref{tab:macro_calibration} in Section \ref{subsec:calibration}. This parametrization is in line with recent papers in the literature of SOE-NK models for Uruguay and/or Latin America.

\begin{table}[ht]
    \centering
    \caption{Model Parametrization \\ Additional Parameters}
    \label{tab:additional_calibration}
    \begin{tabular}{c|c|c}
       Parameter      & Description     & Value \\ \hline \hline
       $\Delta A$     & Steady State Productivity Growth Rate & $\ln(1.03)\times(1/4)$ \\
       $\delta$       & $Z_t$ Moving Average Closeness to $A_t$ & 0.999 \\
       $\eta_g$       & Share of Gov. Exp. in SS. & $0.30$ \\
       $\Upsilon$     & Net Foreign Assets SS.    & $0.4$  \\
       $\theta$       & Calvo Parameter & 0.75 \\
       $\epsilon$     & Final Good Aggregator Elasticity & 6 \\
       $\nu$          & Final Good Firm Subsidy & $1 - (\epsilon-1)/\epsilon$ \\
       $\delta_K$     & Capital Depreciation Rate & 0.025 \\ 
       $\alpha$       & Share of Capital in Production & 0.4 \\
       $\omega^D$     & Share of Domestic Goods in Non-Commodity Aggregator & 0.45 \\
       $\eta_c$       & Elasticity of Substitution Non-Commodity Aggregator & 2.25 \\
       $\gamma_I$     & Share of Domestic Goods in Investment Aggregator & 0.25 \\
       $\nu_I$       & Elasticity of Substitution Investment Aggregator & 0.75 \\
       $\kappa$   & Investment Adjustment Cost & 2 \\
    \end{tabular}
\end{table}

\end{document}